\documentclass[aps,prx,superscriptaddress,twocolumn,showpacs,noeprint]{revtex4-2}

\usepackage{graphicx,amsfonts,times,bm,amsmath,amssymb,verbatim,ifthen,braket,color,array,natbib,bm,subfigure}

\begin{document}

\title{Hybrid quantum gap estimation algorithm using a filtered time series}

\author{Woo-Ram Lee}
\email[Email address:]{wrlee@murrayau.com}
\affiliation{Murray Associates of Utica, Utica, New York 13501, USA}

\author{Ryan Scott}
\affiliation{Department of Physics, Virginia Tech, Blacksburg, Virginia 24061, USA}

\author{V. W. Scarola}
\affiliation{Department of Physics, Virginia Tech, Blacksburg, Virginia 24061, USA}

\begin{abstract}
Quantum simulation advantage over classical memory limitations would allow compact quantum circuits to yield insight into intractable quantum many-body problems, but the interrelated obstacles of large circuit depth in quantum time evolution and noise seem to rule out unbiased quantum simulation in the near term. We prove that classical post-processing, i.e., long-time filtering of an offline time series, exponentially improves the circuit depth needed for quantum time evolution. We apply the filtering method to the construction of a hybrid quantum-classical algorithm to estimate energy gap, an important observable not governed by the variational theorem. We demonstrate, within an operating range of filtering, the success of the algorithm in proof-of-concept simulation for finite-size scaling of a minimal spin model. Our findings set the stage for unbiased quantum simulation to offer memory advantage in the near term.
\end{abstract}

%\pacs{}

\maketitle

\section{Introduction}

Quantum simulation offers the potential not only to speed up solutions to otherwise intractable quantum many-body problems, but it can also yield significant memory advantages in comparison to classical algorithms~\cite{Feynman1982,Lloyd1996,Abrams1999}. Unbiased (exact) classical methods, such as exact diagonalization applied to a time-independent Hamiltonian matrix, $H$, can, in principle, be used to perform finite-size extrapolation of important observables to benchmark approximations, compare with experiment, or map out phase diagrams.  On the other hand, the exponential increase in Hilbert space size of quantum many-body problems severely limits accessible system sizes (i.e., particle or orbital numbers) on classical machines due to memory constraints. The same exponential Hilbert space increase can be leveraged as a memory advantage~\cite{Nielsen2010} by unbiased quantum simulation to compete with classical algorithms on appropriately chosen models~\cite{Yung2014}.  The considerable memory advantage of quantum devices suggests that finite-size quantum simulation could, even in the near term, outperform classical machines in unbiased calculations. 

The quantum phase estimation (QPE)~\cite{Kitaev1996,Lloyd1996,Abrams1999} family of algorithms yields unbiased estimates of various quantities, including energy eigenvalues~\cite{Somma2002,Somma2019,OBrien2019a,Obrien2021,Lin2022,Wang2023,Ding2023a,Ding2023b} and energy gaps~\cite{Wecker2015,Zintchenko2016,Sugisaki2021,Russo2021,Matsuzaki2021}. QPE conventionally relies on the Trotter-Suzuki decomposition~\cite{Trotter1959,Suzuki1976} to implement the time propagator, $e^{-iHt}$, with quantum circuits. Unfortunately, the circuit depth needed to implement time evolution is known to scale rather prohibitively 
% ~\cite{Barthel2020,Childs2021} since long times are needed 
for a speedup advantage~\cite{Barthel2020,Childs2021,Peruzzo2014,McClean2014,Wecker2015a} thus casting doubt on prospects for compact circuit design with QPE.  Furthermore, uncorrected noise in large-depth QPE-based circuits will erode coherence. 

The interrelated obstacles of large circuit depth and noise led to efforts to develop alternative approaches to QPE. Variational quantum eigensolvers (VQEs)~\cite{Peruzzo2014,McClean2016,Tilly2022} turned out to yield noise-tolerant, but biased, estimates of ground-state energy levels. VQE requires only shallow quantum circuits but with sustaining issues, e.g., the Barren plateaus that plague the cost function landscape in scale up~\cite{McClean2018}. Another approach
%, aiming at unbiased quantum simulation, 
starts from the assumption of large numbers of fault-tolerant qubits while designing improvements to scaling of circuit depth. Such ``top-down" approaches have made considerable progress (See Ref.~\cite{Lee2021} for a review), but nonetheless rely on assumptions of high qubit overhead to implement active error correction even on just one single-qubit noise channel~\cite{Tehral2015}. On the other hand, QPE can be revisited from the perspective of hybrid quantum-classical circuits designed for scale up of small noisy quantum devices (``bottom-up approach") for memory advantage (as opposed to speedup advantage). In this approach, as in the case of VQE, classical post-processing is essential but with more focus on unbiased estimates of solutions~\cite{Lin2022,Wang2023,Ding2023a,Ding2023b}.

In this work, we propose a hybrid quantum gap estimation (QGE) algorithm using a filtered time series, built from quantum circuits and post-processed using the classical Fourier transform to return energy gaps. The hybrid QGE algorithm has useful features. First, the quantum circuit  remains compact by avoiding calls to ancilla qubits, quantum Fourier transforms, and quantum state tomography. Second, the solution provides an unbiased estimate of energy gaps beyond the rigorous applicability of the variational theorem. Lastly, the filter in post-processing is used to maximize simulation performance within the operating range. 

Time-series filters, allowing non-unitary evolution of quantum systems, have been implemented in different contexts such as ground-state energy estimation (GSEE) using the approximate cumulative distribution function~\cite{Lin2022,Wang2023} and hybrid dynamical mean-field theory (DMFT) with an impurity solver processed online~\cite{Bauer2016,Keen2020,Steckmann2023}. To our knowledge, however, the role of the filter in gap estimation has not been explored yet. In this work, we show that the filter \emph{exponentially} improves circuit depth at long times.  We demonstrate the performance boost of the hybrid QGE algorithm with time-series filters in proof-of-concept simulations. Noise resilience of the algorithm will be discussed in a separate work~\cite{Lee2024}.

The paper is outlined as follows. In Sec.~\ref{sec_filtered_quantum_time_evolution}, we derive the formula for Trotter truncation error in association with a time-series filter, and reveal the impact on the upper bound of (Trotter) circuit depth. In Sec.~\ref{sec_hybrid_QGE_algorithm}, we describe the hybrid QGE algorithm. In Sec.~\ref{sec_result}, we demonstrate the simulation results for the gap of a minimal spin model and the gap-based phase diagram using finite-size extrapolation. We conclude in Sec.~\ref{sec_conclusion}.

\section{Filtered quantum time evolution}
\label{sec_filtered_quantum_time_evolution}

\subsection{Trotter truncation error with filtering}

QPE-based algorithms leverage the enlarged Hilbert space on quantum devices for evaluation of the time propagator. But intractable Hamiltonians with non-commuting terms, e.g., $H = H_1 + H_2$, where $[H_1,H_2] \neq 0$, are non-trivial to time-evolve. Exact time evolution is described by the time propagator $U_{\rm exact}(t) = e^{-iHt}$. Hereafter, we set $\hbar=1$. The Trotter-Suzuki formula offers various levels of approximation, set by the order $p$, to $U_{\rm exact}(t)$~\cite{Trotter1959,Suzuki1976}: 
\begin{align}
U_M^{(p)}(t) & = [U^{(p)}(t_M)]^M,
\label{Trotter_Suzuki_formula}
\end{align}
where $t_M = t/M$, $M\in\mathbb{N}$, and a single sequence of unitaries is defined by:
\begin{align}
U^{(1)}(t) & = e^{-iH_1 t} e^{-iH_2 t},
\label{Trotter_Suzuki_formula_1st}
\\
U^{(2)}(t) & = e^{-iH_1 t/2} e^{-iH_2 t} e^{-iH_1 t/2},
\label{Trotter_Suzuki_formula_2nd}
\\
U^{(2q)}(t) & = [U^{(2q-2)}(\kappa_{2q} t)]^2 U^{(2q-2)}((1-4\kappa_{2q})t)
\nonumber\\
& ~ \times [U^{(2q-2)}(\kappa_{2q} t)]^2,
\label{Trotter_Suzuki_formula_2pth}
\end{align}
with $\kappa_{2q} = (4-4^{1/(2q-1)})^{-1}$ for $q\geq 2$. Eqs.~\eqref{Trotter_Suzuki_formula}-\eqref{Trotter_Suzuki_formula_2pth} yield precise results once the Trotter depth $M$ exceeds a certain cutoff $M_c$. The $p=1$ formula is especially fit to run on resourced-limited noisy quantum devices. Progress in estimating Trotter truncation error~\cite{Childs2018,Childs2019,Tran2020,Barthel2020,Childs2021} allows us to prove significant improvements in the required cutoff of $M$.

In this section, we prove that long-time filtering of the time propagator leads to substantial improvement in Trotter depth. To start, we take $H\rightarrow H+\Sigma(t)$, where $\Sigma(t)$ defines the self-energy describing energy relaxation to the environment. Here we choose a minimal model~\cite{Haug2007,Comment_selfenergy}: $\Sigma(t) = -i\Gamma(t) I/2$, where $\Gamma(t)>0$ is a user-defined control function to select specific time bins, and $I$ is the identity matrix.

We apply the method in Ref.~\cite{Childs2021} to derive, in the presence of filtering, the leading correction to $U_{\rm exact}(t)$ by Trotter truncation error and thereby the upper bound of $M$. We consider the first-order inhomogeneous differential equation for $\mathcal{U}(t)$: 
\begin{align}
\frac{d}{dt}\mathcal{U}(t) + i\mathcal{H}(t) \mathcal{U}(t) = \mathcal{R}(t),
\label{DiffEq}
\end{align}
where $\mathcal{H}(t)$, $\mathcal{R}(t)$ are continuous operator-valued functions of $t\in\mathbb{R}$. The solution, using variation of parameters, reads as: 
\begin{align}
\mathcal{U}(t) & = Te^{-i\int_0^t d\tau \mathcal{H}(\tau)} \mathcal{U}(0) 
\nonumber\\
& ~ + \int_0^t d\tau_1 Te^{-i\int_{\tau_1}^t d\tau_2 \mathcal{H}(\tau_2)} \mathcal{R}(\tau_1),
\label{solution_general}
\end{align}
where $T$ is the time-ordering operator. To proceed further with $\mathcal{R}(t)$, we can refer back to Eq.~\eqref{DiffEq}. For the application to our problem, we take $\mathcal{U}(t)\rightarrow e^{-\int_0^t d\tau \Gamma(\tau)/2} U_M^{(p)}(t)$, $\mathcal{H}(t)\rightarrow H$ $- i\Gamma(t)I/2$, and factor out the term $e^{-\int_0^t d\tau \Gamma(\tau)/2}$ from both sides of Eq.~\eqref{solution_general}. The equation for $U_M^{(p)}(t)$ is then arranged into the form:
\begin{align}
U_M^{(p)}(t) = U_{\rm exact}(t) + \delta U_M^{(p)}(t),
\label{solution_specific}
\end{align}
where the correction term is given by:
\begin{align}
\delta U_M^{(p)}(t) = \int_0^t d\tau U_{\rm exact}(t-\tau) \big[ \Xi_M^{(p)}(\tau) + \Theta_M^{(p)}(\tau) \big],
\label{U_correction}
\end{align}
with two functions defined in the expansion form:
\begin{align}
\Xi_M^{(p)}(t)
& = \frac{1}{M} \sum_{l=0}^{M-1} [U^{(p)}(t_M)]^{l} \bigg\{\bigg(\frac{d}{dt_M} + iH\bigg) U^{(p)}(t_M)\bigg\} 
\nonumber\\
& ~ \times [U^{(p)}(t_M)]^{M-l-1},
\label{term_Xi}
\\
\Theta_M^{(p)}(t)
& = \bigg\{ iH - \frac{1}{M} \sum_{l=0}^{M-1} [U^{(p)}(t_M)]^{l} iH [U^{(p)}(t_M)]^{-l} \bigg\} 
\nonumber\\
& ~ \times [U^{(p)}(t_M)]^M.
\label{term_Theta}
\end{align}
Eq.~\eqref{term_Xi} gives the leading correction to $U_{\rm exact}(t)$. To proceed, it is convenient to define the adjoint derivative: ${\rm ad}_A B = [A,B]$ along with the related identities: $e^{x{\rm ad}_A}B = e^{xA}Be^{-xA}$, $\frac{d^n}{dx^n}e^{x{\rm ad}_A}B = e^{x{\rm ad}_A}{\rm ad}_A^nB$, where $A$, $B$ are matrices and $x$ is a scalar. For $p=1$, we find:
\begin{align}
\bigg(\frac{d}{dt_M} + iH\bigg) & U^{(1)}(t_M)
= e^{-iH_1t_M} V^{(1)}(t_M) e^{-iH_2t_M},
\label{expansion_form_p1}
\end{align}
where we define
\begin{align}
V^{(1)}(t_M) = (e^{t_M{\rm ad}_{iH_1}} - 1) iH_2.
\label{V1}
\end{align}
Since Eq.~\eqref{V1} satisfies the order condition $\mathcal{O}(t_M)$, it can be represented in the integral form of the Taylor remainder:
\begin{align}
V^{(1)}(t_M) = \int_0^{t_M} d\tau e^{\tau{\rm ad}_{iH_1}} {\rm ad}_{iH_1} iH_2.
\end{align}
Similarly, the result for $p=2$ reads as:
\begin{align}
\bigg(\frac{d}{dt_M} + iH\bigg) U^{(2)}(t_M) 
& = e^{-iH_1t_M/2} V^{(2)}(t_M) 
\nonumber\\
& ~ \times e^{-iH_2t_M} e^{-iH_1t_M/2},
\label{expansion_form_p2}
\end{align}
where we define
\begin{align}
V^{(2)}(t_M) & = - (e^{t_M{\rm ad}_{-iH_2}} - 1) iH_1/2 
\nonumber\\
& ~~~~ + (e^{t_M{\rm ad}_{iH_1/2}} - 1) iH_2.
\label{V2}
\end{align}
Under the order condition $\mathcal{O}(t_M^2)$, Eq.~\eqref{V2} is recast into:
\begin{align}
V^{(2)}(t_M) & = \int_0^{t_M} d\tau \int_0^{\tau} d\tau' 
\big( - e^{\tau'{\rm ad}_{-iH_2}} {\rm ad}_{-iH_2}^2 iH_1/2 
\nonumber\\
& ~ + e^{\tau'{\rm ad}_{iH_1/2}} {\rm ad}_{iH_1/2}^2 iH_2 \big).
\end{align}
For $p\geq 4$, however, complexity increases. At $p=4$, Eq.~\eqref{Trotter_Suzuki_formula_2pth}, for example, has a sequence of 11 unitaries. Here, for our purpose, we just refer to the result in Ref.~\cite{Childs2021}. Lastly, we note that Eq.~\eqref{term_Theta} can be safely ignored because it gives the subleading correction to $U_{\rm exact}(t)$.

In the context of quantum simulation, we consider the trace distance between the exact and Trotterized output states on quantum circuits. In our setup, the output states are represented in the density matrix form:
\begin{align}
\rho_{\rm exact}(t)
& = U_{\rm I}^\dag U_{\rm exact}(t) U_{\rm I} \rho_0 U_{\rm I}^\dag U_{\rm exact}^\dag(t) U_{\rm I},
\label{output_density_matrix_exact}
\\
\rho_M^{(p)}(t)
& = U_{\rm I}^\dag U_M^{(p)}(t) U_{\rm I} \rho_0 U_{\rm I}^\dag [U_M^{(p)}(t)]^\dag U_{\rm I},
\label{output_density_matrix_trotter}
\end{align}
where $U_{\rm I}$ acts on input registers $\rho_0$ for initial state preparation. We plug Eq.~\eqref{solution_specific} in Eq.~\eqref{output_density_matrix_trotter} to find the expansion:
\begin{align}
\rho_M^{(p)}(t) = \rho_{\rm exact}(t) + \delta\rho_M^{(p)}(t) + \mathcal{O}[(\delta\rho_M^{(p)})^2],
\label{output_density_matrix_trotter_expansion}
\end{align}
where the correction term is given by:
\begin{align}
\delta\rho_M^{(p)}(t)
& = U_{\rm I}^\dag \delta U_M^{(p)}(t) U_{\rm I} \rho_0 U_{\rm I}^\dag [U_{\rm exact}(t)]^\dag U_{\rm I}
\nonumber\\
& ~ + U_{\rm I}^\dag U_{\rm exact}(t) U_{\rm I} \rho_0 U_{\rm I}^\dag [\delta U_M^{(p)}(t)]^\dag U_{\rm I}.
\label{output_density_matrix_trotter_1st_order}
\end{align}
In the presence of filtering, we take $\rho(t)\rightarrow\mathcal{F}(t)\rho(t)$, and set Trotter truncation error to: $\varepsilon_{\rm T}^{(p)} \equiv ||\mathcal{F}(t)[\rho_M^{(p)}(t) - \rho_{\rm exact}(t)]||$, where $||A||$ is the spectral norm, i.e., the largest singular value of matrix $A$. $\mathcal{F}(t)$ defines the filter:
\begin{align}
\mathcal{F}(t) = e^{-\int_0^t d\tau \Gamma(\tau)},
\label{filter_function}
\end{align}
which is exemplified by the Lorentzian filter: $\mathcal{F}_{\rm L}(t) = e^{-\eta t}$ for $\Gamma(t) = \eta$, and the Gaussian filter: $\mathcal{F}_{\rm G}(t) = e^{-\sigma^2 t^2/2}$ for $\Gamma(t) = \sigma^2 t$, where $\eta = \sigma \sqrt{2\ln 2}$. To estimate the bound of $\varepsilon_{\rm T}^{(p)}$, we use the properties of the spectral norm: $||cA|| = |c| ||A||$, $||AB|| \leq ||A||~||B||$, $||A+B|| \leq ||A|| + ||B||$, and $||e^{iA}|| = 1$ if $A = A^\dag$, where $A$, $B$ are matrices, and $c$ is a scalar. The multiple integrals are then simplified. For example, $\int_0^t d\tau\int_0^{\tau_M}d\tau' = t^2/(2M)$ for $p=1$, and $\int_0^t d\tau\int_0^{\tau_M}d\tau'$ $\int_0^{\tau'}d\tau'' = t^3/(6M^2)$ for $p=2$. The result is summarized by:
\begin{align}
\varepsilon_{\rm T}^{(p)} 
\leq C^{(p)} \frac{t^{p+1}\mathcal{F}(t)}{M^p},
\label{commutator_scaling}
\end{align}
where the prefactors are defined by:
\begin{align}
C^{(1)} & = ||[H_1,H_2]||,
\label{C_1}
\\
C^{(2)} & = \sum_{\gamma\in\{1,2\}} c_\gamma^{(2)} ||[H_\gamma,[H_1,H_2]]||,
\\
C^{(4)} & = \sum_{\gamma,\lambda,\mu\in\{1,2\}} c_{\gamma\lambda,\mu}^{(4)} ||[H_\gamma,[H_{\lambda},[H_{\mu},[H_1,H_2]]]]||,
\label{C_4}
\end{align}
with numerical constants: 
$c_1^{(2)} = 0.083$, $c_2^{(2)} = 0.167$, 
$c_{111}^{(4)} = 0.0094$, 
$c_{112}^{(4)} = 0.0114$, 
$c_{121}^{(4)} = 0.0092$, 
$c_{122}^{(4)} = 0.0148$, 
$c_{211}^{(4)}$ $= c_{212}^{(4)} = 0.0194$, 
$c_{221}^{(4)} = 0.0346$, 
$c_{222}^{(4)} = 0.0568$. 
We note that, at small times, the upper bound in Eq.~\eqref{commutator_scaling} scales in a different polynomial order of $t$ for each choice of $p$. $\mathcal{F}(t)$ can be used to control the long-time behavior of the bound such that overall suppression is achieved. The role of $\mathcal{F}(t)$ can be addressed more clearly in Fourier space. See Sec.~\ref{sec_postprocessing}. 

Finally, as a consequence of Eq.~\eqref{commutator_scaling}, we find that the total depth for Trotter circuits is bounded above by the cutoff $D_c^{(p)} = N_g^{(p)} M_c^{(p)}$, where $N_g^{(p)}$ counts unitary gates per Trotter iteration, and $M_c^{(p)}$ is Trotter depth cutoff for a fixed $\varepsilon_{\rm T}^{(p)} = \varepsilon_{{\rm T},c}^{(p)}$:
\begin{align}
M_c^{(p)}[\mathcal{F}(t)] 
& = \bigg(\frac{C^{(p)}}{\varepsilon_{{\rm T},c}^{(p)}}\bigg)^{1/p} t^{1+1/p} [\mathcal{F}(t)]^{1/p}.
\label{Trotter_depth_cutoff}
\end{align}
This is one of our central results because it allows us to choose $\mathcal{F}(t)$ to relax otherwise stringent conditions on circuit depth in QPE-based simulation.

\subsection{Example: 1D Transverse-field Ising model}

\begin{figure}[b]
\begin{center}
\includegraphics[width=0.48\textwidth]{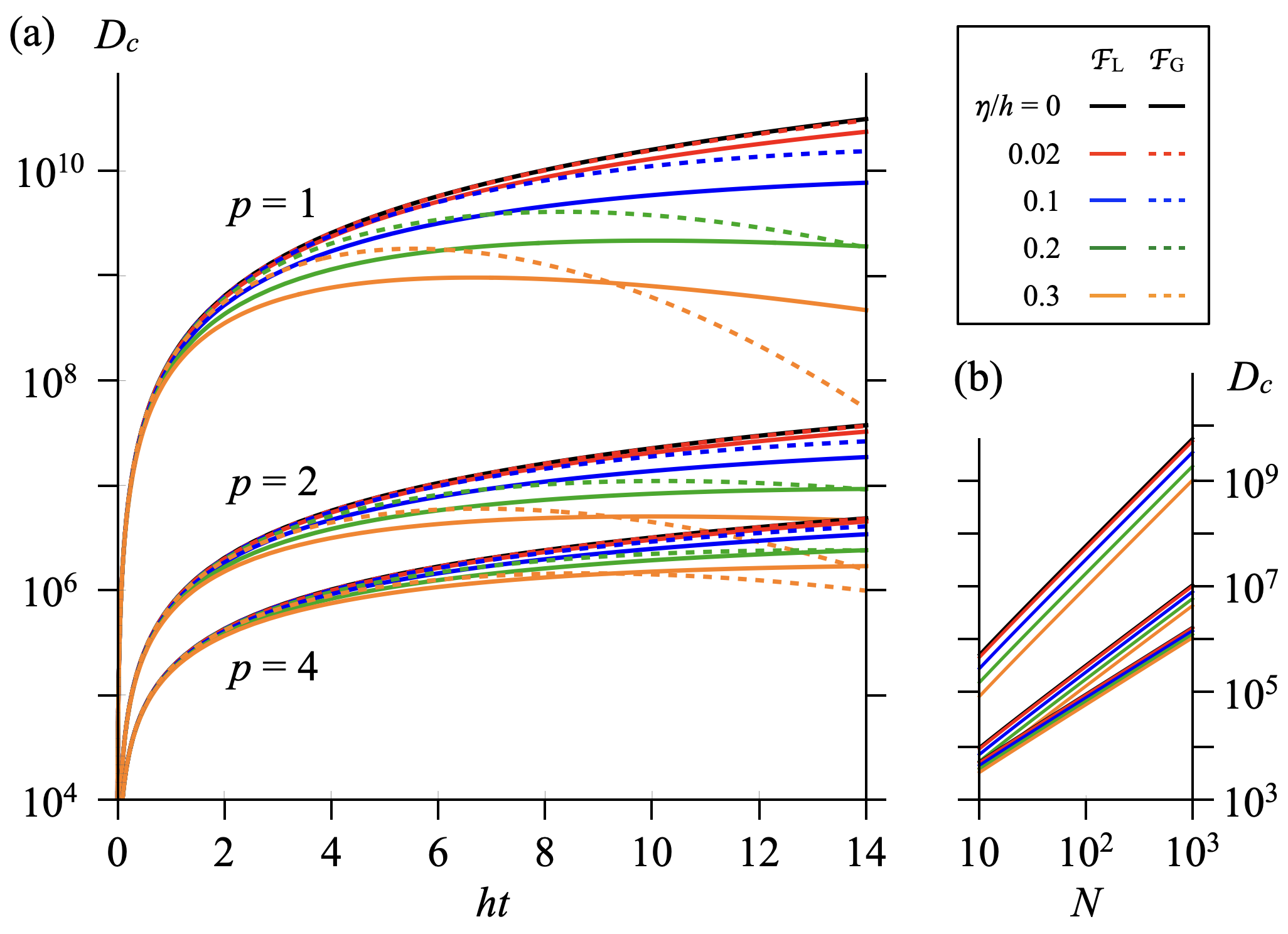}
\end{center}
\caption{(a) Plot of the upper bound of total circuit depth $D_c$ (= Trotter depth cutoff $\times$ gate counts per iteration) versus time for the TFIM ($N = 10^3$) with different choices for the Trotterization order $p\in\{1,2,4\}$ and the filter $\mathcal{F}\in\{\mathcal{F}_{\rm L},\mathcal{F}_{\rm G}\}$, showing that $\eta$ $(=\sigma\sqrt{2\ln 2})$ exponentially suppresses the bound in a long time. Other parameters are set to $J/h = 0.4$, $\varepsilon_{{\rm T},c} = 10^{-2}$. The vertical axis is a log scale. (b) Plot of $D_c$ versus $N$ for a fixed $ht = 6$. Both axes are log scales.}
\label{fig_Trotter_depth}
\end{figure}

To give an estimate to $M_c^{(p)}$, thereby $D_c^{(p)}$, we consider the transverse-field Ising model (TFIM) in one dimension:
\begin{equation}
H_1 = - J \sum_{j=0}^{N-2} \sigma^z_j \sigma^z_{j+1},
~
H_2 = - h \sum_{j=0}^{N-1} \sigma^x_j,
\label{Quantum_Ising_model}
\end{equation}
where $\sigma^{\alpha}$ with $\alpha\in\{x,y,z\}$ are the Pauli matrices, $N$ is the number of spins at sites $j\in[0,N-1]$, $J$ is the Ising coupling, and $h$ is the magnetic field.  The TFIM has a paramagnetic ground state (for $J/h <1$) separated from a ferromagnetic state (for $J/h>1$) by a quantum critical point (at $J/h=1$)~\cite{Sachdev2011}. Straightforward calculations using Eq.~\eqref{Quantum_Ising_model} yield the explicit forms of the commutators in Eqs.~\eqref{C_1}-\eqref{C_4} and the bounds of the spectral norms. See Appendix~\ref{appendix_commutators} for details.

Figure~\ref{fig_Trotter_depth} demonstrates the impact of the filter function on the upper bound of total circuit depth, $D_c$, for the TFIM. Here we count $N_g^{(1)} = N$, $N_g^{(2)} = 2N-1$, $N_g^{(4)} = 6N-1$ (for uncompressed circuits), and use two types of filters $\mathcal{F}_{\rm L}$, $\mathcal{F}_{\rm G}$ for comparison. Without filtering ($\eta=0$), the model shows a prohibitive Trotter scaling that appears to prevent QPE simulation on quantum devices. In Fig.~\ref{fig_Trotter_depth}(a), for spin counts $N=10^3$, top black solid curve ($p=1$) exceeds, for example, $D_c\sim 10^7$ for $ht>6$, which is improved to $D_c\sim 10^4 (10^3)$ for $p=2(4)$. Figure~\ref{fig_Trotter_depth}(b) shows that $|D_c^{(p)} - D_c^{(p'\neq p)}|$ is reduced for smaller $N$. In the presence of filtering ($\eta>0$), however, long-time evolution is truncated and $D_c$ is significantly improved (colored solid curves for $\mathcal{F}_{\rm L}$; colored dashed curves for $\mathcal{F}_{\rm G}$), thus proposing a route to considerable improvements in QPE-based algorithms. Note that filtering is more effective for smaller $p$ [see Eq.~\eqref{Trotter_depth_cutoff}], allowing $D_c^{(p)} > D_c^{(p')}$ for $p<p'$ in a long time. Lastly, $D_c$ is only a bound. In practice, total circuit depth relies on the choice of $H$, $\mathcal{F}$, and the algorithm.

\section{Hybrid quantum gap estimation algorithm}
\label{sec_hybrid_QGE_algorithm}

\subsection{Overview}

\begin{figure}[b]
\begin{center}
\includegraphics[width=0.47\textwidth]{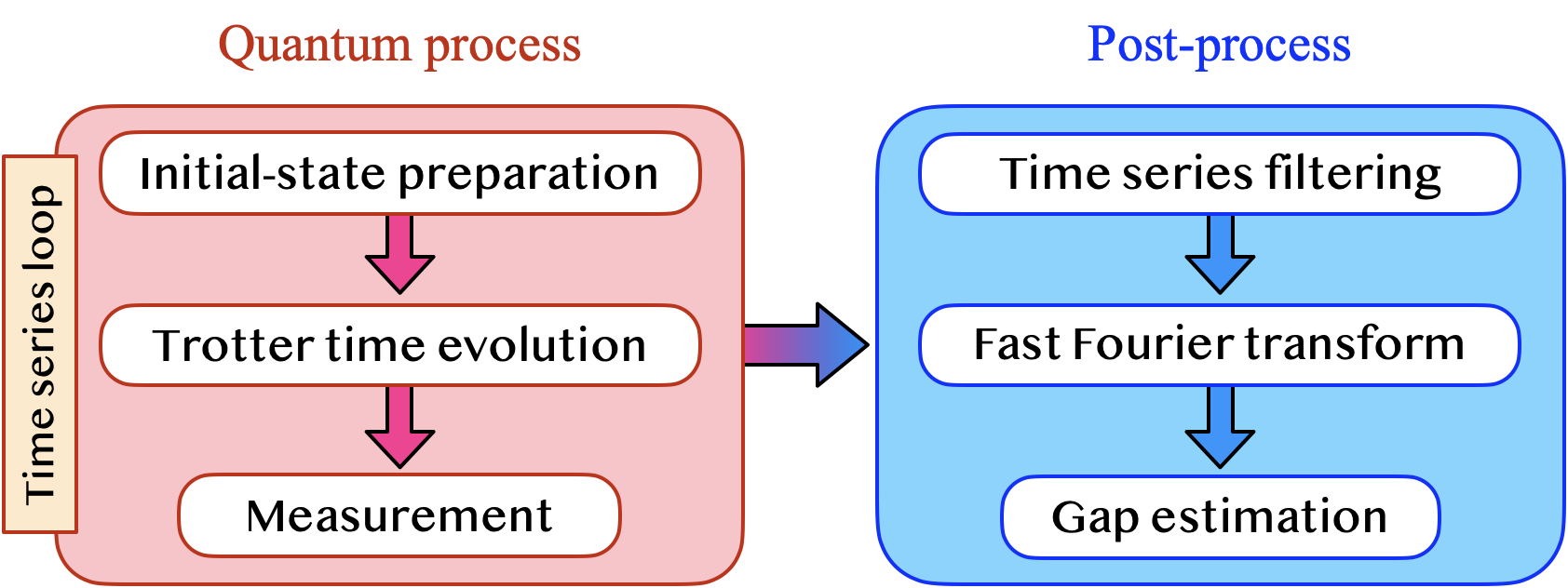}
\end{center}
\caption{
Flowchart for the hybrid QGE algorithm}
\label{fig_flowchart}
\end{figure}

We now construct a hybrid QGE algorithm and and demonstrate the impact of filtering on time evolution. The algorithm flowchart is depicted in Fig.~\ref{fig_flowchart}. We start with an input state $|\psi_{\rm I}\rangle$ that overlaps with an exact state $|\psi_{\rm exact}\rangle$ of interest, perform Trotter time evolution, and then readout in the same basis as the input state. The output state oscillates in time at frequencies of the exact energy gaps for any input state satisfying $\langle\psi_{\rm I}|\psi_{\rm exact}\rangle\neq 0$. A classical fast Fourier transform of a time series, after filtered, reveals exact energy gaps within the window $2\eta$ set by the filter. In the following provided are details for quantum process and post-process, respectively.

\subsection{Quantum process (online)}

\begin{figure}[t]
\begin{center}
\includegraphics[width=0.47\textwidth]{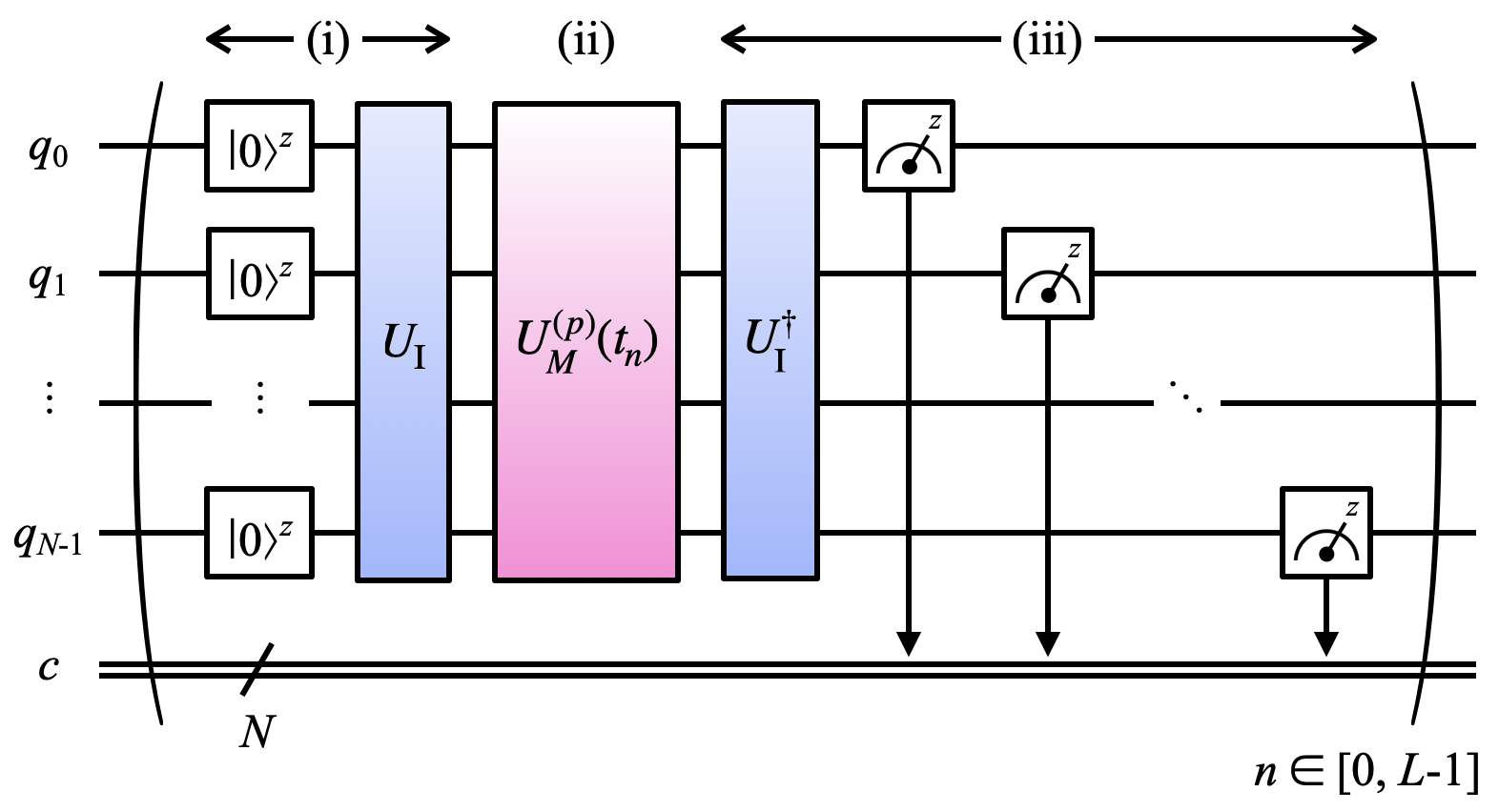}
\end{center}
\caption{Quantum circuit of the hybrid QGE algorithm for a many-body Hamiltonian $H$. (i) Input preparation: $N$ qubits are prepared (and reset) in the quantum registers, ($q_0$, $q_1$, $\cdots$, $q_{N-1}$), and rotated by the unitary $U_{\rm I}$ to create the initial state. (ii) Trotter time evolution of the order $p$ by the unitary $U_M^{(p)}(t_n)$. (iii) Measurement: Output qubits are rotated back to compensate $U_{\rm I}$, and $z$-basis measurements are carried out to return time-series data (of size $L$) to the classical register, $c$.}
\label{fig_quantum_circuit}
\end{figure}

In a quantum processor, each run is iterated over discrete time $t_n = n\delta t$ where $n\in[0,L-1]$ for $L$ Fourier sampling points. Figure~\ref{fig_quantum_circuit} shows the quantum circuit implementation for our example, the TFIM, in a single run. The circuit proceeds in three steps. First, input qubits are prepared in quantum registers to build the initial state: $|\psi_{\rm I}\rangle = U_{\rm I} \prod_{j=0}^{N-1} |0\rangle_j^z$. For our purpose, the product-state unitary is a minimal choice (Here entanglement is not essential):
\begin{align}
U_{\rm I}(\vec{\theta}) = \prod_{j=0}^{N-1} R_j^y(\theta_j),
\end{align} 
where $R_j^\alpha(\theta)=\exp(-i\frac{\theta}{2}\sigma_j^\alpha)$, and $\vec{\theta} = (\theta_0,\theta_1,\cdots,\theta_{N-1})$ is a free parameter that can be chosen to emphasize different gaps. Second, $|\psi_{\rm I}\rangle$ is time-evolved by applying a sequence of unitaries determined by $H$ and $p\in\{1,2,4\}$: 
\begin{align}
& U_M^{(1)}(t_n) = \Big[ \prod_{\vec{j}_2} R_{j_1,j_1+1}^{zz}\Big(\frac{\chi_n}{M}\Big) R_{j_2}^x\Big(\frac{\phi_n}{M}\Big) \Big]^M,
\nonumber\\
& U_M^{(2)}(t_n) = \Big[ \prod_{\vec{j}_3} R_{j_1,j_1+1}^{zz}\Big(\frac{\chi_n}{2M}\Big) R_{j_2}^x\Big(\frac{\phi_n}{M}\Big) R_{j_3,j_3+1}^{zz}\Big(\frac{\chi_n}{2M}\Big) \Big]^M,
\nonumber\\
& U_M^{(4)}(t_n) = \Big[ \prod_{\vec{j}_{11}} R_{j_1,j_1+1}^{zz}\Big(\frac{\kappa_4\chi_n}{2M}\Big) R_{j_2}^x\Big(\frac{\kappa_4\phi_n}{M}\Big) 
\nonumber\\
& ~\times R_{j_3,j_3+1}^{zz}\Big(\frac{\kappa_4\chi_n}{M}\Big) R_{j_4}^x\Big(\frac{\kappa_4\phi_n}{M}\Big) R_{j_5,j_5+1}^{zz}\Big(\frac{(1-3\kappa_4)\chi_n}{2M}\Big) 
\nonumber\\
& ~\times R_{j_6}^x\Big(\frac{(1-4\kappa_4)\phi_n}{M}\Big) R_{j_7,j_7+1}^{zz}\Big(\frac{(1-3\kappa_4)\chi_n}{2M}\Big) R_{j_8}^x\Big(\frac{\kappa_4\phi_n}{M}\Big) 
\nonumber\\
& ~\times R_{j_9,j_9+1}^{zz}\Big(\frac{\kappa_4\chi_n}{M}\Big) R_{j_{10}}^x\Big(\frac{\kappa_4\phi_n}{M}\Big) R_{j_{11},j_{11}+1}^{zz}\Big(\frac{\kappa_4\chi_n}{2M}\Big) \Big]^M,
\end{align}
where $\prod_{\vec{j}_m} = \prod_{j_1} \prod_{j_2}\cdots\prod_{j_m}$, $R_{j,j+1}^{zz}(\theta)$ $= \exp(-i\frac{\theta}{2}\sigma_j^z$ $\sigma_{j+1}^z)$, $\chi_n = -2Jt_n$, $\phi_n = -2ht_n$, and $\kappa_4 = (4-\sqrt[3]{4})^{-1} \approx 0.414$. $M$ repetitions are applied until Trotter error (alternatively, spectral sum or gap estimation error; see Sec.~\ref{sec_convergence_boost}) is reduced below a tolerance. In the last online step, output qubits are rotated back to the input-state basis and measured. Here, quantum state tomography~\cite{Vogel1989} or ancilla qubits~\cite{Somma2002,Sugisaki2021} are not involved, thus saving computational resources.

\subsection{Post-process (offline)}
\label{sec_postprocessing}

The time-evolved output state obtained from the quantum circuit is processed offline. In the first offline step, we build  time-series data, $\{(t_n,\mathcal{P}_n^{(p)})\}_{n=0}^{L-1}$, where measurement outcomes in the $z$ basis are encoded in:
\begin{equation}
\mathcal{P}_{n}^{(p)} 
= {\rm Tr}[\rho_0\rho_{\vec{\theta}}^{(p)}(t_n)].
\label{propagator}
\end{equation}
Here we define the density matrices $\rho_0 =$ $\prod_{j=0}^{N-1} |0\rangle_j^z \langle 0|_j^z$ for the input registers and $\rho_{\vec{\theta}}^{(p)}(t_n) = U_{M,\vec{\theta}}^{(p)}(t_n) \rho_0 [U_{M,\vec{\theta}}^{(p)}(t_n)]^\dag$ for the output with the similarity transform $U_{M,\vec{\theta}}^{(p)}(t_n) = [U_{\rm I}(\vec{\theta})]^\dag$ $U_M^{(p)}(t_n) U_{\rm I}(\vec{\theta})$. Next, Eq.~\eqref{propagator} is filtered by $\mathcal{F}_n = \mathcal{F}(t_n)$, and fed into the classical subroutine for a discrete Fourier transform (DFT) which yields a (many-body) spectral function:
\begin{equation}
\mathcal{A}^{(p)} (\omega_m) = \frac{\delta t}{2\pi} {\rm Re} \sum_{s=\pm} \sum_{n=0}^{L-1} e^{i\omega_m t_{sn}} \mathcal{F}_n \mathcal{P}_{sn}^{(p)},
\label{spectral_function}
\end{equation}
where we define discrete frequencies $\omega_m = m\delta\omega$, conjugate to $t_n$, in units of $\delta t$ and $\delta\omega$ satisfying $\delta\omega \delta t = 2\pi/L$, and $m,n \in [0,L-1]$. The $s=+(-)$ terms in Eq.~\eqref{spectral_function} describe causal (anti-causal) processes. In practice, a fast Fourier transform (FFT) is widely adopted to improve computational complexity of the original DFT, $\mathcal{O}(N^2)$ to $\mathcal{O}(N\log N)$~\cite{Duhamel1990}.

To reveal key features of Eq.~\eqref{spectral_function}, we consider the continuum limit ($L\rightarrow\infty$):
\begin{equation}
\mathcal{A}^{(p)}(\omega) 
= \frac{1}{2\pi} \sum_{s=\pm} \int_{-\infty}^\infty dt \cos(\omega t) \Theta(st) \mathcal{F}(st) \mathcal{P}^{(p)}(t),
\label{spectral_function_continuum}
\end{equation}
where $\Theta(t)$ is the step function, and $\mathcal{P}^{(p)}(t) = {\rm Tr}[\rho_0\rho_{\vec{\theta}}^{(p)}(t)]$ $= |\langle\psi_{\rm I}|U_M^{(p)}(t)|\psi_{\rm I}\rangle|^2\in\mathbb{R}$, that is distinguished from the form $\langle\psi_{\rm I}|U_M^{(p)}(t)|\psi_{\rm I}\rangle\in\mathbb{C}$ for eigenvalue estimation~\cite{Comment_QEE}. Eq.~\eqref{spectral_function_continuum} can be recast further in the convolution form: 
\begin{equation}
\mathcal{A}^{(p)}(\omega) = \int_{-\infty}^\infty d\tilde{\omega}\tilde{\mathcal{F}}(\tilde{\omega})\mathcal{A}_0^{(p)}(\omega-\tilde{\omega}),
\label{spectral_function_convolution}
\end{equation}
where $\tilde{\mathcal{F}}$ represents the filter in Fourier space:
\begin{equation}
\tilde{\mathcal{F}}(\omega) 
= \frac{1}{\pi}\int_{0}^\infty dt \cos(\omega t) \mathcal{F}(t),
\label{filter_transformed}
\end{equation}
and $\mathcal{A}_0^{(p)}(\omega) \equiv \mathcal{A}^{(p)}(\omega)|_{\mathcal{F}=1}$ (without filtering). Two features of Eq.~\eqref{spectral_function_convolution} are addressed below.

First, peak centers in $\mathcal{A}_0^{(p)}$ return exact energy gaps for any choices of $H$ and $|\psi_{\rm I}\rangle$. To show this, ignoring Trotter error for $M>M_c$, we approximate $U_M^{(p)}(t)\approx e^{-iHt}$. We then expand $|\psi_{\rm I}\rangle = \sum_u c_u|u\rangle$, where eigenstates $|u\rangle$ satisfy $H|u\rangle = \mathcal{E}_u|u\rangle$ with eigenenergies $\mathcal{E}_u$, and plug it in $\mathcal{A}_0^{(p)}$ to derive the spectral representation (Hereafter, we drop the superscript, $p$):
\begin{equation}
\mathcal{A}_0(\omega) = \sum_{u,v} |c_u|^2 |c_v|^2 \delta(\omega - \Delta_{u,v}),
\label{spectral_function_decomposed}
\end{equation}
where we define the Dirac delta function $\delta(\omega) = \frac{1}{2\pi}\int dt e^{i\omega t}$ and exact energy gaps $\Delta_{u,v} = \mathcal{E}_u - \mathcal{E}_v$. For estimation of a target, say, $\Delta_{u,v}$, the condition of $|c_u||c_v|\neq 0$ is generally demanded. Without loss of generality, we can drop the redundant sum over $u\leq v$ to focus on the partial sum over $u>v$.

Second, Eq.~\eqref{spectral_function_convolution} converts delta functions in Eq.~\eqref{spectral_function_decomposed} into the line shapes set by Eq.~\eqref{filter_transformed}:
\begin{equation}
\mathcal{A}(\omega) = \sum_{u,v} |c_u|^2 |c_v|^2 \tilde{\mathcal{F}}(\omega - \Delta_{u,v}).
\label{spectral_function_decomposed_filtered}
\end{equation}
For the choice of $\mathcal{F}_{\rm L}(t) = e^{-\eta t}$, $\tilde{\mathcal{F}}(\omega)$ describes a Lorentzian line shape: $\tilde{\mathcal{F}}_{\rm L}(\omega) = \frac{1}{\pi}\frac{\eta}{\omega^2 + \eta^2}$. For $\mathcal{F}_{\rm G}(t) = e^{-\sigma^2 t^2/2}$, it turns out to be a Gaussian line shape: $\tilde{\mathcal{F}}_{\rm G}(\omega) = \frac{1}{\sqrt{2\pi}\sigma} e^{-\frac{\omega^2}{2\sigma^2}}$. Here, the broadening $2\eta$ defines the full width at half maximum of the line shape, and has a connection to $\sigma$: $\eta = \sigma\sqrt{2\ln 2}$. It has an operating range designed to maximize simulation performance: $2\eta$ is bounded below by Trotter error $\varepsilon_{\rm T}$ and above by peak-to-peak separations to hold spectral resolutions in gap estimation. See Sec.~\ref{sec_operating_range_filtering} for further discussion.

The last offline step establishes a consistent gap estimation protocol for Eq.~\eqref{spectral_function}. We need to start with an initial guess of the energy gap, $\Delta_0$, e.g., mean-field or perturbative.  Here we use perturbation theory for the TFIM with open boundaries: $\Delta_0/h = 2[1-(1-1/N)J/h]$ (See Appendix~\ref{appendix_energy_gap_formula}). (We focus on the lowest gap but can find any gap by adjusting the initial guess.) We then search for the peak center in the range $\Delta_0-\delta\Delta/2\leq\omega\leq\Delta_0+\delta\Delta/2$ to find the unbiased estimate of $\Delta$. Here the search window $\delta\Delta$ is initially set to $2\eta$. If $\Delta$ is not within the range, we restart with either a new choice of $\delta\Delta$ or $\Delta_0$. This process is iterated until a solution is found.

\section{Results}
\label{sec_result}

So far, we have proposed the hybrid QGE algorithm using a filtered time series, thereby requiring shallow circuits. In this section, we demonstrate proof-of-concept simulation results using the TFIM of the small size $N\in\{2,3,4,5\}$ as a benchmark model~\cite{Comment_GitHub}. The results are compared between different choices of filter $\mathcal{F}$ and Trotterization order $p$. Here the study on noise-induced error in quantum processes is beyond our scope. To demonstrate the algorithm, we discuss our implementation case: gaps of the TFIM and gap-based paramagnetic phase diagram determined by finite-size extrapolation.

\begin{figure*}[t]
\begin{center}
\includegraphics[width=0.98\textwidth]{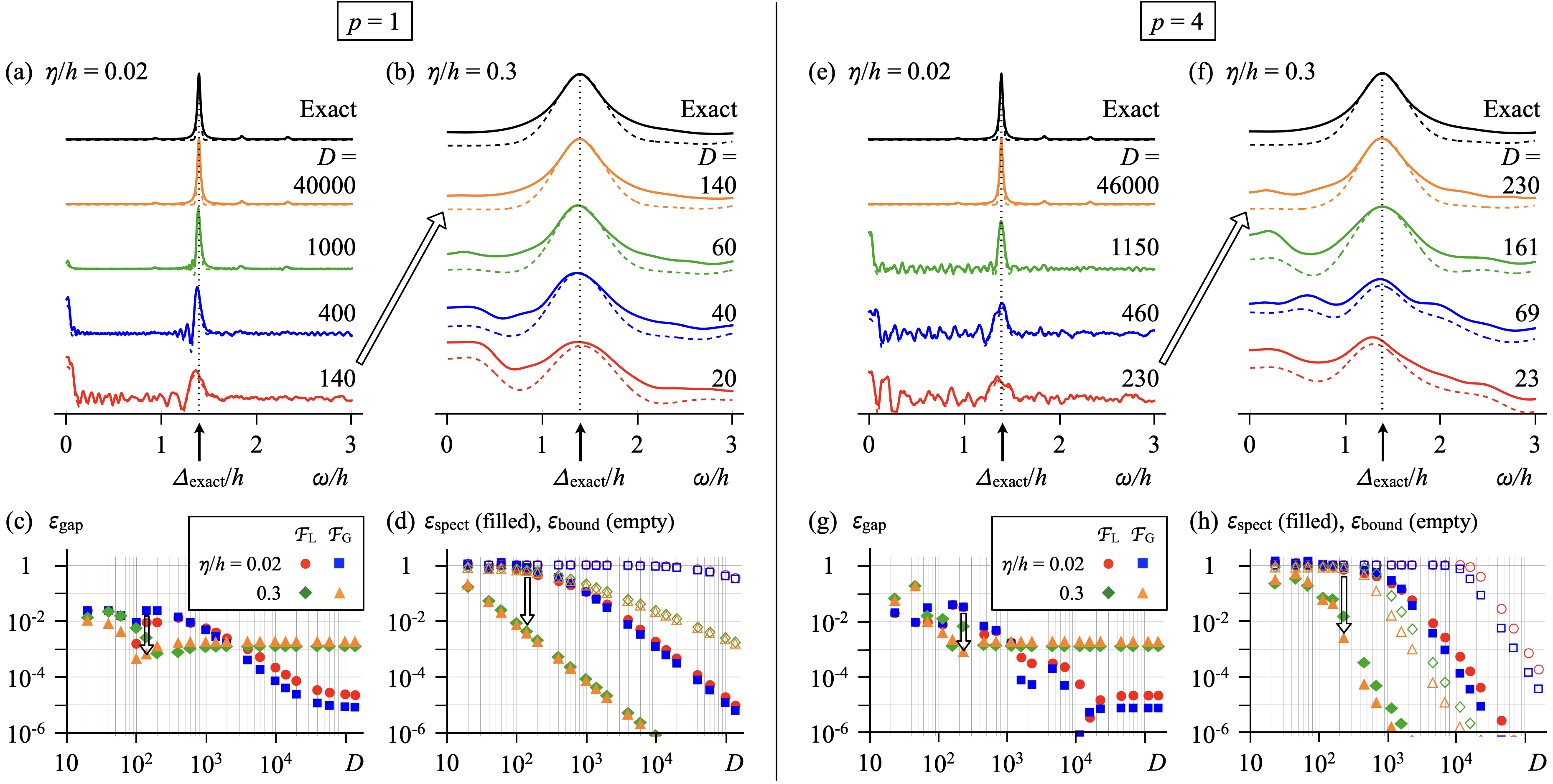}
\end{center}
\caption{Simulation results demonstrating convergence boost of QGE by filtering. The results are compared between different choices of the Trotterization order $p=$ (a-d) 1, (e-h) 4, the broadening $\eta/h =$ (a,e) 0.02, (b,f) 0.3, and the filter $\mathcal{F}\in\{\mathcal{F}_{\rm L},\mathcal{F}_{\rm G}\}$. The top panels show the spectral function $\mathcal{A}$ versus frequency $\omega$ for the TFIM of $N=4$, $J/h = 0.4$. Colored solid (dashed) curves are the results for $\mathcal{F} = \mathcal{F}_{\rm L}$ ($\mathcal{F}_{\rm G}$). All cases show convergence to the exact form (top black curve) for increasing circuit depth $D$. The upper bound of $D$ is effectively reduced by increasing $\eta$. Black vertical dotted line refers to the exact gap $\Delta_{\rm exact}$. Tilted black empty arrow shows how $\mathcal{A}$ evolves as $\eta$ increases for $D =$ (a,b) 140, (e,f) 230. The bottom panels indicate (c,g) gap estimate error $\varepsilon_{\rm gap}$ and (d,h) spectral lineshape error $\varepsilon_{\rm spect}$ (filled symbols), bounded above by $\varepsilon_{\rm bound}$ (empty symbols), as a function of $D$ for different choices of $\mathcal{F}$, $\eta$. In all simulations, 1024 measurement shots were used, and other parameters are: $\theta_j = 0.27\pi$ (uniform over sites), $\delta\omega = \eta/4$, $L = 2\lceil 7h/\delta\omega\rceil$~\cite{Comment_parameters}, where $\lceil x\rceil$ is the ceiling function of $x$.}
\label{fig_convergence}
\end{figure*}

\subsection{Convergence boost of QGE by filtering}
\label{sec_convergence_boost}

The central assertion in Sec.~\ref{sec_filtered_quantum_time_evolution} was that filtering Trotter time evolution at long times effectively lowers the upper bound of circuit depth for a fixed Trotter truncation error. Here we numerically confirm that the filtering method can be leveraged to boost the convergence of the hybrid QGE algorithm. 

Figure~\ref{fig_convergence} shows simulation results for $N=4$, $J/h=0.4$ but with different choices of $p\in\{1,4\}$, $\mathcal{F}\in\{\mathcal{F}_{\rm L},\mathcal{F}_{\rm G}\}$. Input orientations are fixed here, but, later on, will be tuned for further investigation. The top panels indicate that increasing $\eta$ gives better convergence of $\mathcal{A}$ to the exact form with $D$. The plots with $D\ll \tilde{D}_c$, where $\tilde{D}_c$ is the empirical bound of circuit depth, are featured by satellite peaks (apart from the main peak) which are governed by the Floquet stroboscopic dynamics~\cite{Heyl2019}. If the satellite peaks are separated within the resolution limit $2\eta$, they are smoothed out, thereby leaving only the main peak behind. Black empty arrow highlights $D=140$ $(230)$ for $p=1(4)$, allowing direct comparison between the plots with various $\eta$ but fixed $D$. 

Either choice of $p$ or $\mathcal{F}$ can change the characteristics of convergence while overall trends hold. To show this clearly, we measure different types of errors from the data and compare them. First off, we define the gap estimate error:
\begin{align}
\varepsilon_{\rm gap} =  \frac{|\Delta - \Delta_{\rm exact}|}{\Delta_{\rm exact}},
\label{gap_estimate_error}
\end{align}
where $\Delta$ is a gap estimate obtained using the protocol outlined in Sec.~\ref{sec_postprocessing}, and $\Delta_{\rm exact}$ is the exact gap as a reference for a small-size system. Figure~\ref{fig_convergence}(c),(g) show the evolution of $\varepsilon_{\rm gap}$ with $D$. For each choice of ($\mathcal{F}$, $\eta$), we can estimate the circuit depth cutoff $\tilde{D}_c$ where $\varepsilon_{\rm gap}$ reaches the lower bound. Plateaus arise since the frequency resolution is limited by $2\eta$. We find that $\tilde{D}_c$ is lowered for larger $\eta$ with an extra shift relying on the choice of $\mathcal{F}$: $\tilde{D}_{c,{\rm G}} < \tilde{D}_{c,{\rm L}}$. Such a distinction is because the Gaussian line shape is more concentrated at the center than the Lorentzian case, efficiently suppressing interference with neighbor peaks. Meanwhile, the effect of higher $p$ is not as impressive as we might expect. Comparing the $p=4$ data with $p=1$, for example, $\tilde{D}_c$ is slightly lowered (lifted) for $\eta/h=0.02(0.3)$. In fact, this is nothing but what Fig.~\ref{fig_Trotter_depth} implies: For $\eta\rightarrow 0$, $p=4$ outperforms $p=1$ in the entire time domain, while that is not the case for large $\eta$ since $p=1$ is improved more than $p=4$ at long times. 

\begin{figure*}[t]
\begin{center}
\includegraphics[width=0.985\textwidth]{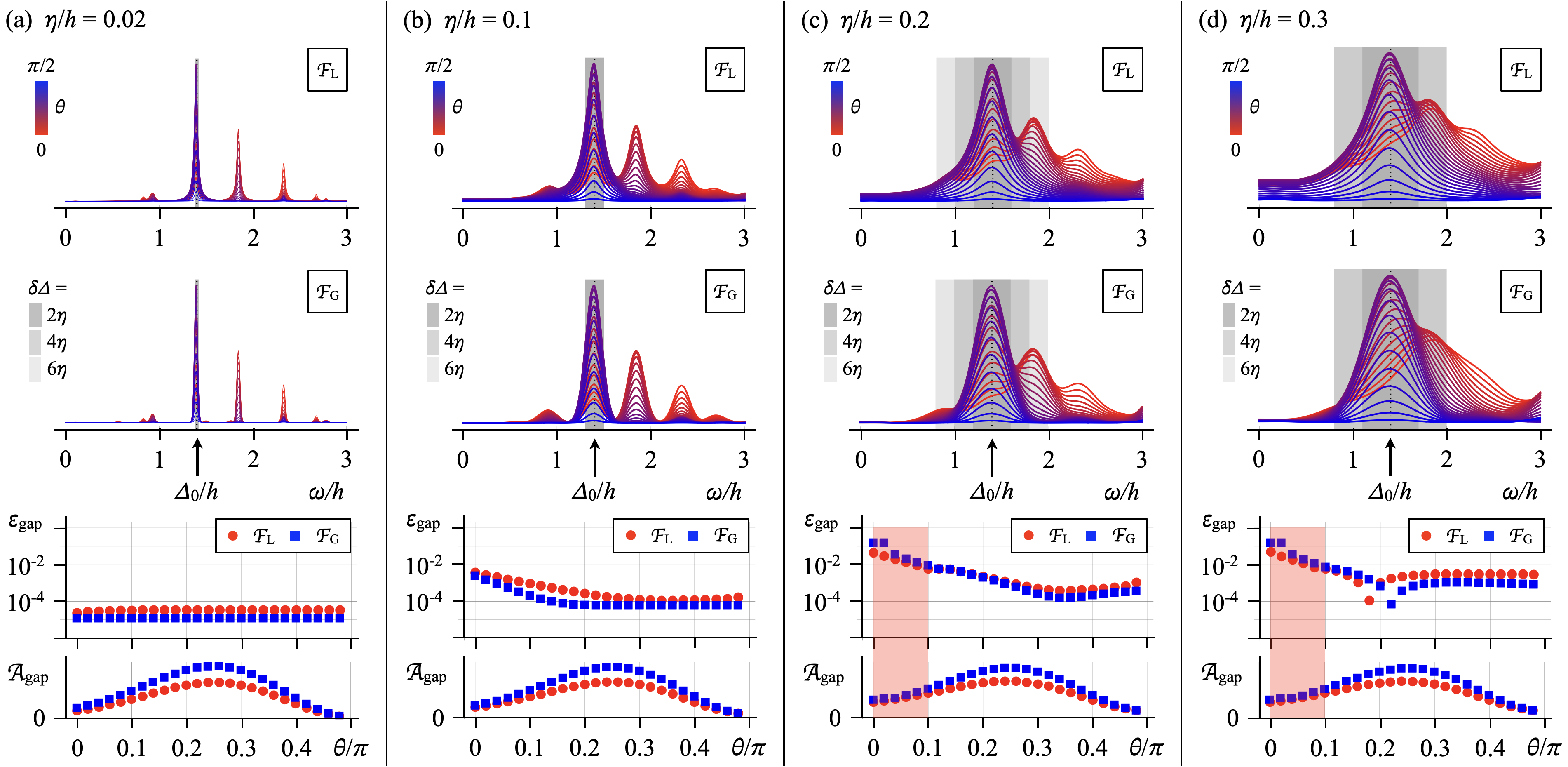}
\end{center}
\caption{Simulation results demonstrating the operating range of filtering. The results are compared between different choices of the broadening $\eta/h=$ (a) 0.02, (b) 0.1, (c) 0.2, (d) 0.3, the filter $\mathcal{F}\in\{\mathcal{F}_{\rm L},\mathcal{F}_{\rm G}\}$, but with Trotterization set to the order $p=1$. The top and middle panels show the spectral function $\mathcal{A}$ versus frequency $\omega$ for the TFIM of $N=4$, $J/h = 0.4$ but with different $\mathcal{F}$. Here circuit depth is set to $D=$ (a) 40000, (b) 1400, (c) 400, (d) 140 (satisfying the spectral lineshape error $\varepsilon_{\rm spect}<10^{-3}$), and $\mathcal{A}$'s are sampled with uniformly selected input orientations $\theta_l/\pi = l/50$, $l\in[0,24]$ (but irrespective of sites). For each $\theta_l$, starting with the initial guess for energy gap $\Delta_0$ (black vertical dotted line), we search for the peak center close to $\Delta_0$ within the window $[\Delta_0-\delta\Delta/2,\Delta_0+\delta\Delta/2]$ by progressively increasing $\delta\Delta$ (gray shades). The bottom panels show the gap estimate error $\varepsilon_{\rm gap}$ and the peak height $\mathcal{A}_{\rm gap}=\mathcal{A}(\omega=\Delta)$, respectively, as a function of $\theta$ for both $\mathcal{F}_{\rm L},\mathcal{F}_{\rm G}$. Red shades indicate the unfavored zone where we set $\varepsilon_{\rm gap}\gtrsim 10^{-2}$. Other parameters are the same as Fig.~\ref{fig_convergence}.}
\label{fig_input_orientation}
\end{figure*}

Another useful measure is the spectral lineshape error (as a square root of the measure known as the coefficient of determination in statistics~\cite{Chicco2021}):
\begin{align}
\varepsilon_{\rm spect} = \sqrt{\frac{\sum_{m=0}^{L-1}[\mathcal{A}(\omega_m) - \mathcal{A}_{\rm exact}(\omega_m)]^2}{\sum_{m=0}^{L-1}[\mathcal{A}(\omega_m) - \mathcal{A}_{\rm av}]^2}},
\label{spectral_lineshape_error}
\end{align}
where $\mathcal{A}_{\rm av} = \frac{1}{L}\sum_{m=0}^{L-1}\mathcal{A}(\omega_m)$. Since Eq.~\eqref{gap_estimate_error} only measures error in the main peak center, error may be accidentally reduced when satellite peaks dwell around the main peak for small $D$. Eq.~\eqref{spectral_lineshape_error}, by contrast, accumulates errors in the spectral line shape over the entire frequency domain, and therefore provides a more consistent measure. Figures~\ref{fig_convergence}(d),(h) show the counterpart to Figs.~\ref{fig_convergence}(c),(g) for different choices of $(p,\mathcal{F},\eta)$. The overall trend of $\varepsilon_{\rm spect}$ is matched with $\varepsilon_{\rm gap}$, but without plateaus since the resolution of $\varepsilon_{\rm spect}$ is not affected by $\eta$. 

Lastly, it is useful to derive the upper bound of Eq.~\eqref{spectral_lineshape_error}. A starting point is the Trotter truncation error $||\mathcal{F}(t_n)[\rho_M(t_n) - \rho_{\rm exact}(t_n)]||$ with the upper bound [Eq.~\eqref{commutator_scaling}]. Since the spectral norm $||\cdot||$ plays a key role in deriving the upper bound, we define $\hat{\mathcal{A}}(\omega_m)$ by replacing ${\rm Tr}[\cdot]$ in Eq.~\eqref{spectral_function} by $||\cdot||$. Then the upper bound of Eq.~\eqref{spectral_lineshape_error} has the form:
\begin{equation}
\varepsilon_{\rm bound} 
= \sqrt{\frac{\sum_{m=0}^{L-1}[\hat{\mathcal{A}}(\omega_m) - \hat{\mathcal{A}}_{\rm exact}(\omega_m)]^2}{\sum_{m=0}^{L-1}[\hat{\mathcal{A}}(\omega_m) - \hat{\mathcal{A}}_{\rm av}]^2}},
\label{spectral_lineshape_error_bound}
\end{equation}
where $\hat{\mathcal{A}}(\omega_m) = \hat{\mathcal{A}}_{\rm exact}(\omega_m) + \delta\hat{\mathcal{A}}(\omega_m)$, with
\begin{equation}
\left\{
\begin{array}{c}
\hat{\mathcal{A}}_{\rm exact}
\\
\delta\hat{\mathcal{A}}
\end{array}
\right\}
= \frac{\delta t}{2\pi} {\rm Re} \sum_{s=\pm} \sum_{n=0}^{L-1} e^{i\omega_m t_{sn}} \mathcal{F}_n \left\{
\begin{array}{c}
1
\\
C^{(p)} t_{sn}^{p+1} / M^p
\end{array}
\right\},
\end{equation}
and $\hat{\mathcal{A}}_{\rm av} = \frac{1}{L}\sum_{m=0}^{L-1} \hat{\mathcal{A}}(\omega_m)$. Empty symbols in Fig.~\ref{fig_convergence}(d),(h) represent Eq.~\eqref{spectral_lineshape_error_bound}, revealing a similar trend to filled symbols but with overestimation in all range of circuit depth.

\subsection{Operating range of filtering}
\label{sec_operating_range_filtering}

We just showed that the filtering method can be leveraged to improve the convergence of the hybrid QGE algorithm but only with the input unitary $U_{\rm I}(\vec{\theta})$ fixed to a certain form. In fact, $U_{\rm I}(\vec{\theta})$ provides a control knob that allows further investigation of our algorithm. Importantly, the filter has an operating range designed to maximize simulation performance. Specifically, $\eta$ in the filter is bounded above by spectral resolution set by the algorithm and model. Tuning $\vec{\theta}$ in $U_{\rm I}(\vec{\theta})$ can impact the resolution by changing relative peak heights. Here we map out the spectral functions with different ratios of peak heights for different choices of $\vec{\theta}$ to numerically confirm the upper bound of $\eta$.

\begin{figure*}[t]
\begin{center}
\includegraphics[width=0.98\textwidth]{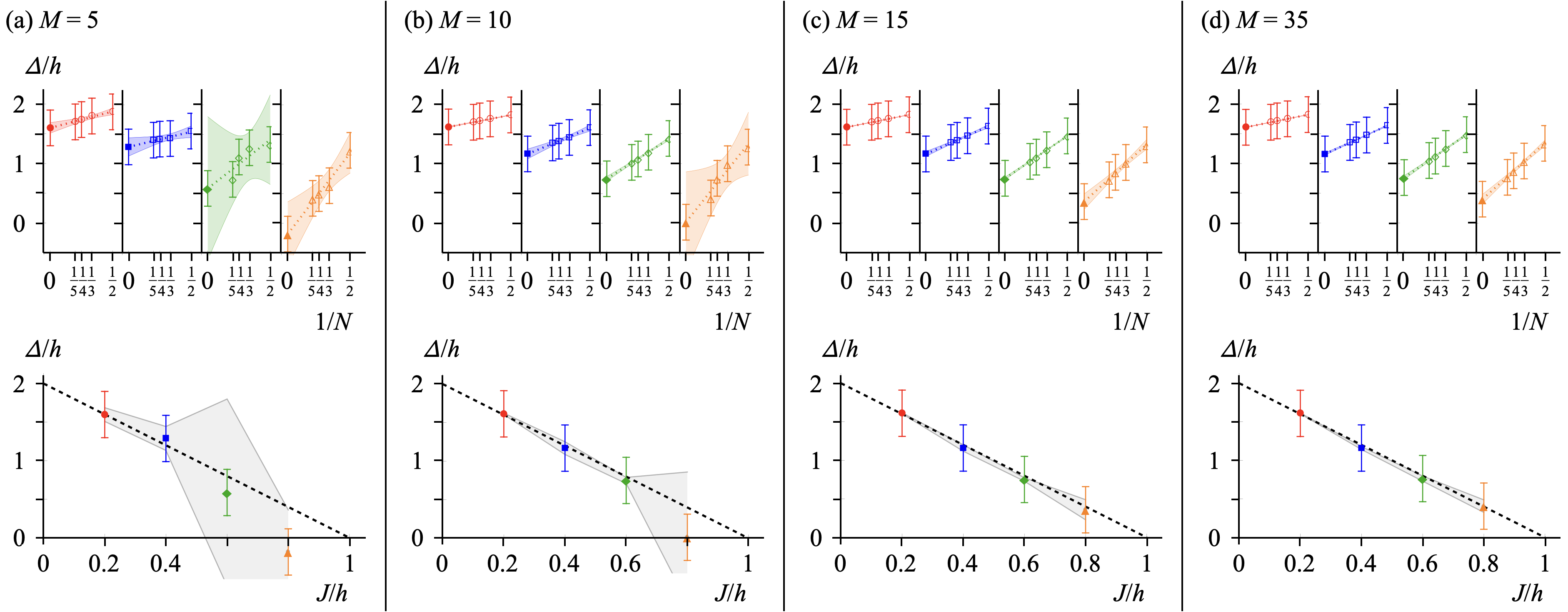}
\end{center}
\caption{Stages of construction for the gap-based phase diagram of quantum paramagnet. In the hybrid QGE simulation, we fix the Trotterization order $p=1$, the filter $\mathcal{F} = \mathcal{F}_{\rm G}$, and the broadening $\eta/h = 0.3$. Each column shows progress on convergence for increasing Trotter depth $M =$ (a) 5, (b) 10, (c) 15, (d) 35. The top panels show finite-size scaling of energy gap for the TFIM. Empty symbols indicate the samples of gap estimates $\{\Delta_{N,J/h}\}$ for $N=2-5$, $J/h =$ 0.2 (red), 0.4 (blue), 0.6 (green), 0.8 (orange). Vertical fence symbols set error bounds $[\Delta-\eta,\Delta+\eta]$ around $\Delta$. Filled symbols are extrapolation to $N\rightarrow\infty$ by linear regression (dotted lines). Colored shades are confidence bands with a 95\% confidence level. The bottom panels show the gap-based phase diagram of quantum paramagnet ($J<h$). Filled symbols are the extrapolated data points at $N\rightarrow\infty$ (obtained from the top panels). Gray shades are interpolation of the confidence band edges (at $N\rightarrow\infty$) for different $J/h$. Black dashed line indicates the exact gap $\Delta_{\rm exact}^{(N\rightarrow\infty)}$. Other parameters are the same as Fig.~\ref{fig_convergence}.}
\label{fig_finite-size_scaling}
\end{figure*}

Figure~\ref{fig_input_orientation} shows simulation results for $N=4$, $J/h=0.4$ but with different choices of $\eta/h\in\{0.02,0.1,0.2,0.3\}$. Here we use $p=1$ Trotterization. The top and middle panels, assuming $\theta_j = \theta$, i.e., uniform input orientation over sites $j$, indicates that tuning $\theta\in[0,\pi/2]$ emphasizes different peaks in $\mathcal{A}$ for the choice of $\mathcal{F}_{\rm L}$ (top), $\mathcal{F}_{\rm G}$ (middle). The gap is estimated starting with an initial guess $\Delta_0$ (indicated by arrow) and progressively increasing the search window $\delta\Delta$ (gray shades). The bottom panel shows the gap estimate error $\varepsilon_{\rm gap}$ as a function of $\theta$. According to Eq.~\eqref{spectral_function_decomposed}, a gap estimate without filtering is invariant under any choice of $\theta$. This is exemplified by Fig.~\ref{fig_input_orientation}(a), where $2\eta$ is far below peak-to-peak resolution, and thus $\varepsilon_{\rm gap}$ is constant over $\theta$. Meanwhile, filtering makes each peak broadened by $2\eta$ [see Eq.~\eqref{spectral_function_decomposed_filtered}], thus allowing the overlap between neighbor peaks. The tail of neighbor peaks generally forms a slanted background that shifts the peak center in our interest. Figure~\ref{fig_input_orientation}(b-d), in contrast to Fig.~\ref{fig_input_orientation}(a), reveals a non-uniform modification to $\varepsilon_{\rm gap}$ over $\theta$, growing for increasing $\eta$. Specifically, the modification actively arises in the regime of $\theta$ close to 0, where the neighbor peak centered at $\omega/h\approx 1.845$ grows and eventually dominates the main peak. For larger $\eta$, a minor peak is buried more easily under the background, thus shifting the peak center from one to another and lifting $\varepsilon_{\rm gap}$. The choice of $\mathcal{F}$ also affects the above argument since it sets the background in a different form. In Fig.~\ref{fig_input_orientation}(d), for example, $\mathcal{F}_{\rm G}$ more effectively lowers $\varepsilon_{\rm gap}$ than $\mathcal{F}_{\rm L}$ for $\theta\geq 0.2\pi$, while it is reversed for $\theta<0.2\pi$. As mentioned before, this is attributed to the fact that the Gaussian line shape is more concentrated at the center than the Lorentzian case. Appendix~\ref{appendix_model_spectral_function} describes a toy model to support our argument on peak center shift. Lastly, the plots for $\mathcal{A}_{\rm gap}$, estimated at $\omega=\Delta$, show that one can safely avoid the unfavored zone in $\theta$ [red shades in Fig.~\ref{fig_input_orientation}(c,d)] by maximizing the peak height. As discussed later, this condition can serve as a protocol to construct a phase diagram.

\subsection{Finite-size scaling of quantum paramagnetic gap}
\label{sec_finite-size_scaling}

Previously, we demonstrated the performance of the hybrid QGE algorithm using shallow quantum circuits in the filtering method, and discussed its validity. Our algorithm can be combined with post-processes to calculate important physical quantities such as a (gap-based) phase diagram. As a benchmark, we construct the phase diagram of quantum paramagnet ($J<h$), and compare the simulation result with the exact solution. Here we focus on the case with $p=1$, $\mathcal{F} = \mathcal{F}_{\rm G}$, and $\eta/h = 0.3$.

Construction is organized into three stages. First, from the simulation, we obtain the sample of gap estimates $\{\Delta_{N,J/h}\}$ for $N\in\{2,3,4,5\}$, $J/h\in\{0.2,0.4,0.6,0.8\}$. For $\eta$ comparable to spectral resolution [See Fig.~\ref{fig_input_orientation}(c,d)], we need a protocol for addressing the range of $\theta$ where $\varepsilon_{\rm gap}$ stays minimal. As discussed before, a sufficient condition is to maximize the height of a target peak. Empty symbols in the top panels of Fig.~\ref{fig_finite-size_scaling} indicate data sampled in the above manner. Next, for each choice of $J/h$, we use linear regression to extrapolate the data for $N\in\{2,3,4,5\}$ to $N\rightarrow\infty$ (filled symbols). The result is accompanied by confidence bands (colored shades) that represent misalignment of data points. In the last stage, the extrapolated data is rearranged to construct the phase diagram of quantum paramagnet. The bottom panels in Fig.~\ref{fig_finite-size_scaling} show the results. Here the black dashed lines compare with the exact gap $\Delta_{\rm exact}^{(N\rightarrow\infty)} = 2|h-J|$ (See Appendix~\ref{appendix_energy_gap_formula}). 

Finally, we confirm convergence of the phase diagram for increasing Trotter depth $M$. The columns in Fig.~\ref{fig_finite-size_scaling} are arranged in the ascending order of $M$ [in the same way as Fig.~\ref{fig_convergence}(b)]. For small $M$, $\Delta$ is off from $\Delta_{\rm exact}$ and confidence bands are comparable to/exceed error bounds $2\eta$ (depending on the choice of $J/h$), while, for increasing $M$, they are safely getting inside $2\eta$ and eventually converge to $\Delta_{\rm exact}$. 

\section{Conclusion}
\label{sec_conclusion}

We developed a hybrid QGE algorithm using a filtered time series. We found, using the filter, exponential improvement of circuit depth at long times, and mapped out the role of various input states to reveal the operating range of filtering. We finally showed how our protocol can be used. We constructed the gap-based paramagnetic phase diagram for a minimal spin model, which demonstrates how a quantum device can offer memory advantage in finite-size extrapolation of energy gaps.

Further improvements by, e.g., Cartan decomposition~\cite{Kokcu2022} and Bayesian methods~\cite{Wiebe2016,Zintchenko2016,Sugisaki2021}, could allow applications to many-body models requiring more gates to implement time evolution. Our approach can also be applied to hybrid DMFT algorithms~\cite{Bauer2016,Keen2020,Steckmann2023}, where speedup and noise resilience were recently observed~\cite{Steckmann2023}, and a recent proposal of measurement-based hybrid algorithm for eigenvalue estimation~\cite{Lee2022}.

\begin{acknowledgments}
We acknowledge support from AFOSR (FA2386-21-1-4081, FA9550-19-1-0272, FA9550-23-1-0034) and ARO (W911NF2210247, W911NF2010013). We thank A.F. Kemper and P. Roushan for insightful discussion. We acknowledge the use of IBM Quantum services for this work. The views expressed are those of the authors, and do not reflect the official policy or position of IBM or the IBM Quantum team.
\end{acknowledgments}

\appendix

\section{Explicit forms of Eqs.~\eqref{C_1}-\eqref{C_4} for the TFIM}
\label{appendix_commutators}

Here, using Eq.~\eqref{Quantum_Ising_model} for the TFIM, we explicitly calculate the spectral norms in Eqs.~\eqref{C_1}-\eqref{C_4}. The commutators of $H_1$ and $H_2$ of interest can be successively derived as follows:
\begin{align}
& [H_1, H_2] 
= 2iJh \sum_{j=0}^{N-2} (\sigma^y_j \sigma^z_{j+1} + \sigma^z_j \sigma^y_{j+1}),
\label{Commutator_12}
\\
& [H_1,[H_1, H_2]] 
= -8J^2h \bigg[\sum_{j=0}^{N-1} \sigma^x_j + \sum_{j=0}^{N-3} \sigma^z_j \sigma^x_{j+1} \sigma^z_{j+2}\bigg],
\label{commmutator_112}
\\
& [H_2,[H_1, H_2]]
= -8Jh^2 \sum_{j=0}^{N-2} (\sigma^y_j \sigma^y_{j+1} - \sigma^z_j \sigma^z_{j+1}),
\label{commmutator_212}
\\
& [H_1,[H_1,[H_2,[H_1, H_2]]]] = [H_1,[H_2,[H_1,[H_1, H_2]]]] 
\nonumber\\
& = -64J^3h^2 \bigg[ \sum_{j=0}^{N-2} \sigma^y_j \sigma^y_{j+1} - \sum_{j=0}^{N-4} \sigma^z_j \sigma^x_{j+1} \sigma^x_{j+2} \sigma^z_{j+3} \bigg],
\\
& [H_2,[H_1,[H_2,[H_1, H_2]]]] = [H_2,[H_2,[H_1,[H_1, H_2]]]] 
\nonumber\\
& = 64J^2h^3 \sum_{j=0}^{N-3} (\sigma^y_j \sigma^x_{j+1} \sigma^y_{j+2} - \sigma^z_j \sigma^x_{j+1} \sigma^z_{j+2}).
\end{align}
All others are connected with Eqs.~\eqref{commmutator_112} and \eqref{commmutator_212}:
\begin{align}
& [H_\gamma,[H_1,[H_1,[H_1, H_2]]]] = 8J^2 [H_\gamma,[H_1, H_2]],
\\
& [H_\gamma,[H_2,[H_2,[H_1, H_2]]]] = 16h^2 [H_\gamma,[H_1, H_2]],
\label{Commutator_21212}
\end{align}
where $\gamma\in\{1,2\}$. In the above derivation, we used the identities: $[AB,C] = A[B,C]$ $+ [A,C]B$ for the matrices $A$, $B$, $C$, $[\sigma^\alpha, \sigma^\beta] = 2i\varepsilon_{\alpha\beta\gamma}\sigma^\gamma$, $\{\sigma^\alpha, \sigma^\beta\} = 2\delta_{\alpha\beta} I$ with the Kronecker delta $\delta_{\alpha\beta}$, the Levi-Civita symbol $\varepsilon_{\alpha\beta\gamma}$, and $\alpha,\beta,\gamma\in\{x,y,z\}$. For the spectral norms of Eqs.~\eqref{Commutator_12}-\eqref{Commutator_21212}, we apply the properties: $||cA|| = |c| ||A||$, $||AB|| \leq ||A||~||B||$, $||A+B|| \leq ||A|| + ||B||$, and $||e^{iA}|| = 1$ if $A = A^\dag$, where $A$, $B$ are matrices, and $c$ is a scalar. It turns out that the upper bounds of the commutators are summarized:
\begin{align}
||[\tilde{H}_1, \tilde{H}_2]|| 
& \leq 4(N - 1),
\label{spectral_norm_2}
\\
||[\tilde{H}_\gamma,[\tilde{H}_1, \tilde{H}_2]]||
& \leq 16(N-1),
\\
||[\tilde{H}_\gamma,[\tilde{H}_\lambda,[\tilde{H}_{\bar{\lambda}},[\tilde{H}_1, \tilde{H}_2]]]]||
& \leq 128(N-2),
\\
||[\tilde{H}_\gamma,[\tilde{H}_\lambda,[\tilde{H}_\lambda,[\tilde{H}_1, \tilde{H}_2]]]]||
& \leq 128\lambda(N-1),
\label{spectral_norm_5}
\end{align}
where $\tilde{H}_1 = H_1/|J|$, $\tilde{H}_2 = H_2/|h|$, $\gamma,\lambda\in\{1,2\}$, and $\bar{\lambda}=1(2)$ if $\lambda=2(1)$. Eqs.~\eqref{spectral_norm_2}-\eqref{spectral_norm_5} allows deriving the upper bounds of Eqs.~\eqref{C_1}-\eqref{C_4}, thereby Eq.~\eqref{commutator_scaling}.

\section{Reference formulas for energy gap of the TFIM}
\label{appendix_energy_gap_formula}

Here we derive the reference formulas for energy gap of the TFIM using (i) perturbation theory, (ii) exact methods.

\subsection{Initial guess of energy gap}

\begin{figure}[b]
\begin{center}
\includegraphics[width=0.48\textwidth]{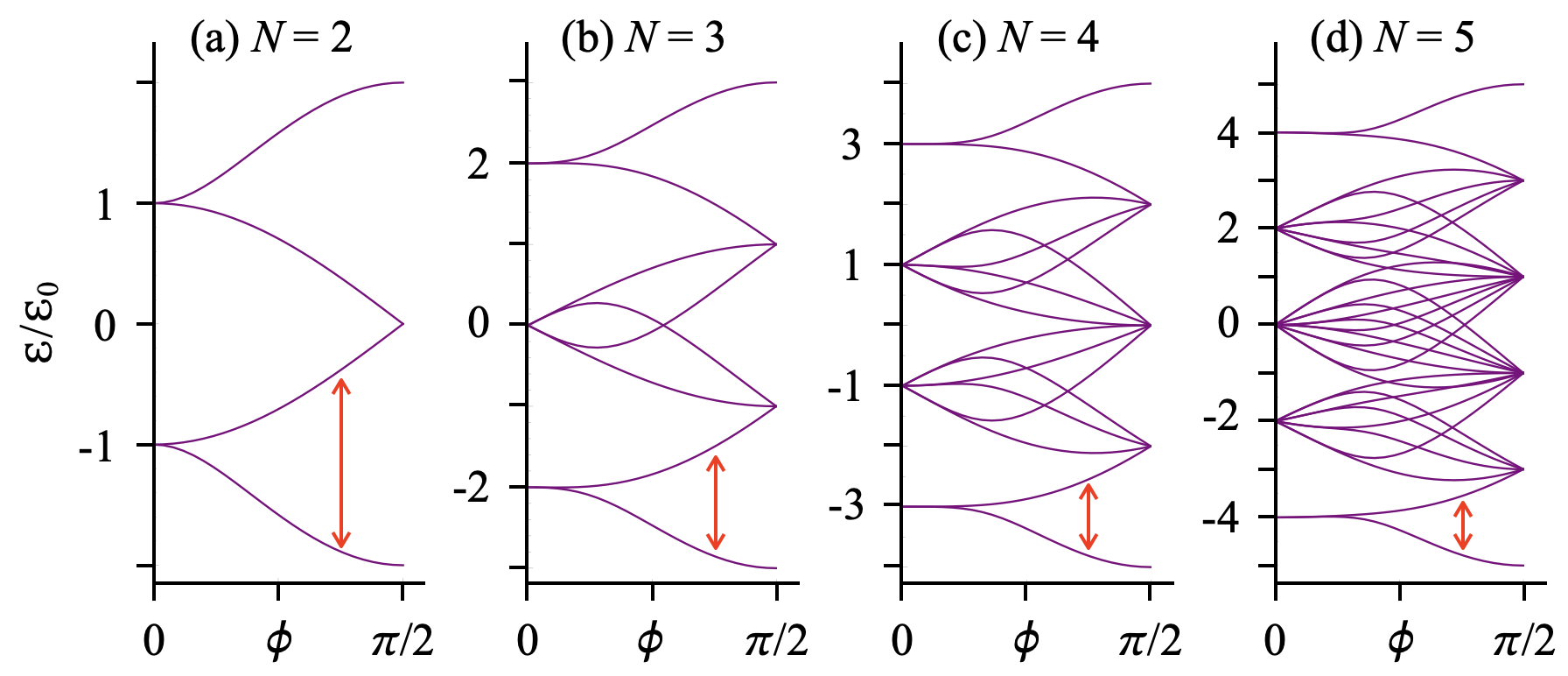}
\end{center}
\caption{Energy spectrum of the TFIM for $N=2-5$ as a function of $\phi[=\cot^{-1}(J/h)]$. Here we have ferromagnet (or paramagnet) at $\phi = 0$ (or $\pi/2$), and red arrow indicates the lowest energy gap of quantum paramagnet ($\pi/4 < \phi \leq \pi/2$).
}
\label{fig_energy_spectrum}
\end{figure}

Figure~\ref{fig_energy_spectrum} shows the energy spectrum of the TFIM for $N=2-5$, where the lowest energy gap of interest is indicated by the red arrow. In the hybrid QGE algorithm, gap estimation generally requires an initial guess of the energy gap. For this purpose, we consider a generally applicable procedure: perturbation expansion~\cite{Sachdev2011}. In our case, we perturb in powers of $J/h$ to find the approximate energy gap of the quantum paramagnet. For the TFIM with $N$ spins, a spin-flip from the paramagnetic ground state $\prod_{j=1}^N|0\rangle_j^z$ requires excitation energy $2h - 2J$ for $N-2$ bulk spins and $2h - J$ for two boundary spins. Averaging excitation energies over all spins yields an approximate formula for energy gap:
\begin{align}
\Delta_0/h & = [(N-2)(2h-2J) + 2(2h-J)]/(Nh) 
\nonumber\\
& = 2[1 - (1-1/N)J/h],
\end{align}
that is reduced to the case with periodic boundaries~\cite{Sachdev2011}:
\begin{equation}
\Delta_0/h|_{N\rightarrow\infty} = 2(1 - J/h).
\label{perturbation_gap_periodic}
\end{equation}
This expression for $\Delta_0$ shows how we derived the initial guess for the gap and also establishes that perturbative methods can, in other models, be used to define the guess. 

\subsection{Exact energy gap}

The TFIM is tractable in the limit $N\rightarrow\infty$. An exact solution provides a reference to compare with the simulation result. Using the Jordan-Wigner transformation~\cite{Jordan1928}: $\sigma_j^x = 1 - $ $2c_j^\dag c_j$, $\sigma_j^z = -\prod_{k<j}(1-2c_k^\dag c_k)(c_j + c_j^\dag)$, the TFIM is mapped to the Kitaev model describing p-wave superconductor:
\begin{equation}
H_{\rm K} = - w\sum_{j=1}^{N-1} (c_j^\dag c_{j+1} + c_j^\dag c_{j+1}^\dag + {\rm H.c.}) - \mu \sum_{j=1}^N \delta n_j,
\end{equation}
where $w(=J)$ is the hopping/pairing energy, $\mu(=-2h)$ is the chemical potential, and $\delta n_j = c_j^\dag c_j^{\vphantom{\dagger}}$ $- 1/2$. Assuming periodic boundaries, $H_{\rm K}$ can be diagonalized by the Bogoliubov transformation~\cite{Sachdev2011}. The result yields the exact form of the upper and lower energy bands:
\begin{align}
\mathcal{E}_{\bf k}^\pm = \pm\sqrt{J^2 + h^2 - 2Jh\cos(ka)},
\end{align}
where $k$ is momentum, and $a$ is the lattice constant. The lowest energy gap between two bands is therefore given by the energy difference at ${\bf k} = 0$:
\begin{equation}
\Delta_{\rm exact}^{(N\rightarrow\infty)} = \mathcal{E}_{{\bf k}=0}^+ - \mathcal{E}_{{\bf k}=0}^- = 2|h - J|,
\end{equation}
consistent with Eq.~\eqref{perturbation_gap_periodic} in the paramagnetic regime ($J<h$).

\section{Toy model for peak center shift}
\label{appendix_model_spectral_function}

\begin{figure}[b]
\begin{center}
\includegraphics[width=0.49\textwidth]{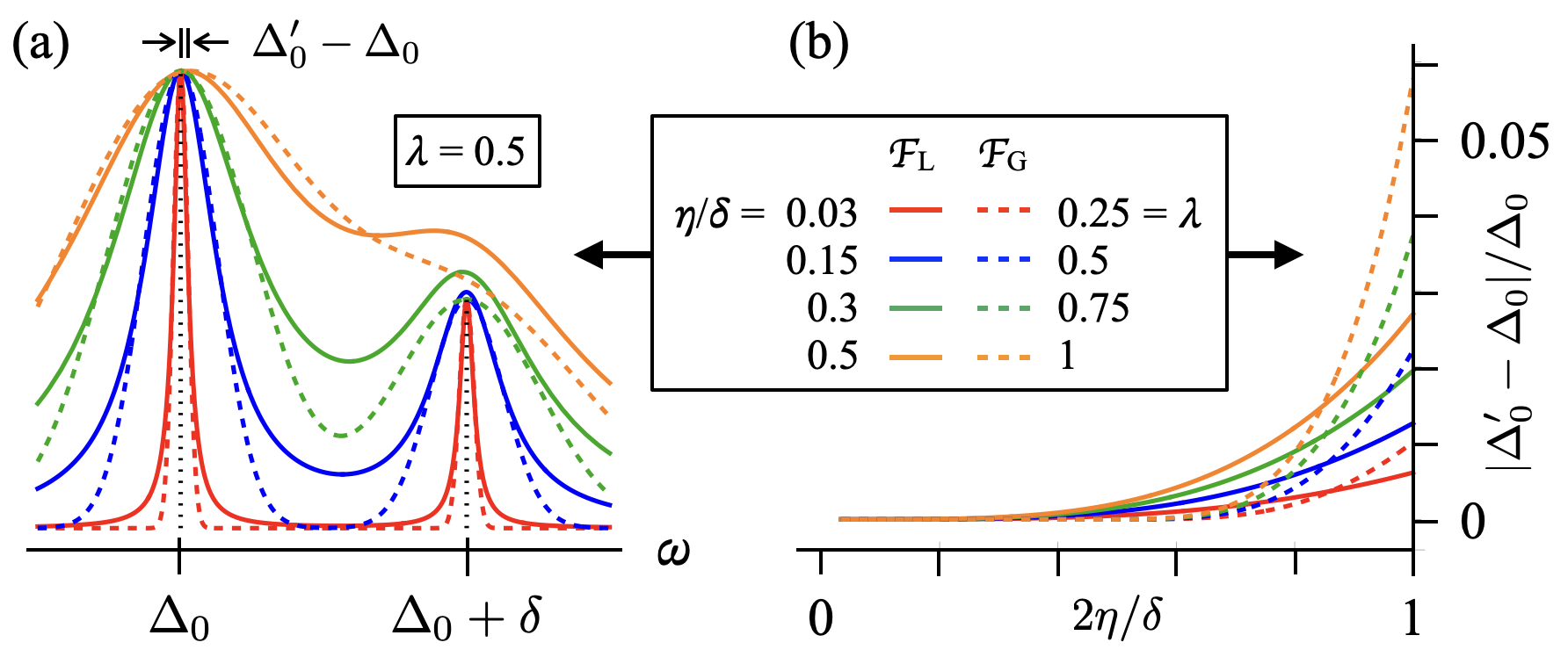}
\end{center}
\caption{(a) Spectral function for two peaks separated by $\delta/\Delta_0 = 0.6$ with different choices of the broadening $\eta$ but fixed relative height $\lambda = 0.5$. Here all plots are normalized and the vertical black dotted lines indicate the reference to the peak centers at $\omega = \Delta_0,\Delta_0+\delta$ for $\eta\rightarrow 0$. (b) Peak center shift $|\Delta'_0 - \Delta_0|/\Delta_0$ as a function of $\eta$ for different choices of $\lambda$ and the filter $\mathcal{F}\in\{\mathcal{F}_{\rm L},\mathcal{F}_{\rm G}\}$.}
\label{fig_peak_center_shift}
\end{figure}

Here we describe a toy model for peak center shift to support the argument in Sec.~\ref{sec_operating_range_filtering}. In general, peak centers are subject to shift when the separation between peaks is comparable to the broadening of each peak. To show this, we model the spectral function for two peaks separated by $\delta$:
\begin{equation}
\mathcal{A}(\omega) \propto \mathcal{A}_0(\omega - \Delta_0) + \lambda \mathcal{A}_0(\omega - \Delta_0 - \delta),
\label{two_peak_spectral_function}
\end{equation}
where we set 
\begin{align}
\mathcal{A}_0(\omega) = \left\{
\begin{array}{cc}
\frac{1}{\pi}\frac{\eta}{\omega^2 + \eta^2}, & \textrm{Lorentzian}
\\
\frac{1}{\sqrt{2\pi}\sigma} e^{-\frac{\omega^2}{2\sigma^2}}, & \textrm{Gaussian}
\end{array}
\right.
\end{align}
describing an isolated Lorentzian (Gaussian) peak associated with the filter $\mathcal{F}_{\rm L(G)}$, $\eta = \sigma\sqrt{2\ln 2}$, and $\lambda$ controls the second peak height. Figure~\ref{fig_peak_center_shift}(a) shows how much the original line shape is modified when two peaks are overlapped in the tail for different choices of $\eta$ but fixed $\lambda$. In Fig.~\ref{fig_peak_center_shift}(b), we focus on the first peak at $\omega = \Delta_0$ and measure the peak center shift $|\Delta'_0 - \Delta_0|/\Delta_0$ as a function of $\eta$ for different choices of $\lambda$. The result implies: (i) the peak center shift is enhanced for increasing $\eta$, (ii) it is enhanced (suppressed) for increasing (decreasing) symmetry between peaks, (iii) $\mathcal{F}_{\rm L(G)}$ induces larger shift than $\mathcal{F}_{\rm G(L)}$ for $2\eta/\delta\lesssim(\gtrsim)$ 0.85.

\bibliography{references} 

%Control: production of eprint (1) enabled
\begin{thebibliography}{52}%
\makeatletter
\providecommand \@ifxundefined [1]{%
 \@ifx{#1\undefined}
}%
\providecommand \@ifnum [1]{%
 \ifnum #1\expandafter \@firstoftwo
 \else \expandafter \@secondoftwo
 \fi
}%
\providecommand \@ifx [1]{%
 \ifx #1\expandafter \@firstoftwo
 \else \expandafter \@secondoftwo
 \fi
}%
\providecommand \natexlab [1]{#1}%
\providecommand \enquote  [1]{``#1''}%
\providecommand \bibnamefont  [1]{#1}%
\providecommand \bibfnamefont [1]{#1}%
\providecommand \citenamefont [1]{#1}%
\providecommand \href@noop [0]{\@secondoftwo}%
\providecommand \href [0]{\begingroup \@sanitize@url \@href}%
\providecommand \@href[1]{\@@startlink{#1}\@@href}%
\providecommand \@@href[1]{\endgroup#1\@@endlink}%
\providecommand \@sanitize@url [0]{\catcode `\\12\catcode `\$12\catcode
  `\&12\catcode `\#12\catcode `\^12\catcode `\_12\catcode `\%12\relax}%
\providecommand \@@startlink[1]{}%
\providecommand \@@endlink[0]{}%
\providecommand \url  [0]{\begingroup\@sanitize@url \@url }%
\providecommand \@url [1]{\endgroup\@href {#1}{\urlprefix }}%
\providecommand \urlprefix  [0]{URL }%
\providecommand \Eprint [0]{\href }%
\providecommand \doibase [0]{https://doi.org/}%
\providecommand \selectlanguage [0]{\@gobble}%
\providecommand \bibinfo  [0]{\@secondoftwo}%
\providecommand \bibfield  [0]{\@secondoftwo}%
\providecommand \translation [1]{[#1]}%
\providecommand \BibitemOpen [0]{}%
\providecommand \bibitemStop [0]{}%
\providecommand \bibitemNoStop [0]{.\EOS\space}%
\providecommand \EOS [0]{\spacefactor3000\relax}%
\providecommand \BibitemShut  [1]{\csname bibitem#1\endcsname}%
\let\auto@bib@innerbib\@empty
%</preamble>
\bibitem [{\citenamefont {Feynman}(1982)}]{Feynman1982}%
  \BibitemOpen
  \bibfield  {author} {\bibinfo {author} {\bibfnamefont {R.~P.}\ \bibnamefont
  {Feynman}},\ }\bibfield  {title} {\bibinfo {title} {Simulating physics with
  computers},\ }\href {https://doi.org/10.1007/BF02650179} {\bibfield
  {journal} {\bibinfo  {journal} {Int. J. Theor. Phys.}\ }\textbf {\bibinfo
  {volume} {21}},\ \bibinfo {pages} {467} (\bibinfo {year} {1982})}\BibitemShut
  {NoStop}%
\bibitem [{\citenamefont {Lloyd}(1996)}]{Lloyd1996}%
  \BibitemOpen
  \bibfield  {author} {\bibinfo {author} {\bibfnamefont {S.}~\bibnamefont
  {Lloyd}},\ }\bibfield  {title} {\bibinfo {title} {Universal quantum
  simulators},\ }\href {https://doi.org/10.1126/science.273.5278.1073}
  {\bibfield  {journal} {\bibinfo  {journal} {Science}\ }\textbf {\bibinfo
  {volume} {273}},\ \bibinfo {pages} {1073} (\bibinfo {year}
  {1996})}\BibitemShut {NoStop}%
\bibitem [{\citenamefont {Abrams}\ and\ \citenamefont
  {Lloyd}(1999)}]{Abrams1999}%
  \BibitemOpen
  \bibfield  {author} {\bibinfo {author} {\bibfnamefont {D.~S.}\ \bibnamefont
  {Abrams}}\ and\ \bibinfo {author} {\bibfnamefont {S.}~\bibnamefont {Lloyd}},\
  }\bibfield  {title} {\bibinfo {title} {{Quantum Algorithm Providing
  Exponential Speed Increase for Finding Eigenvalues and Eigenvectors}},\
  }\href {https://doi.org/10.1103/PhysRevLett.83.5162} {\bibfield  {journal}
  {\bibinfo  {journal} {Phys. Rev. Lett.}\ }\textbf {\bibinfo {volume} {83}},\
  \bibinfo {pages} {5162} (\bibinfo {year} {1999})}\BibitemShut {NoStop}%
\bibitem [{\citenamefont {Nielsen}\ and\ \citenamefont
  {Chuang}(2010)}]{Nielsen2010}%
  \BibitemOpen
  \bibfield  {author} {\bibinfo {author} {\bibfnamefont {M.~A.}\ \bibnamefont
  {Nielsen}}\ and\ \bibinfo {author} {\bibfnamefont {I.~L.}\ \bibnamefont
  {Chuang}},\ }\href {https://books.google.com/books?id=-s4DEy7o-a0C} {\emph
  {\bibinfo {title} {\textit{Quantum Computation and Quantum Information}}}}\
  (\bibinfo  {publisher} {Cambridge University Press},\ \bibinfo {year}
  {2010})\BibitemShut {NoStop}%
\bibitem [{\citenamefont {Yung}\ \emph {et~al.}(2014)\citenamefont {Yung},
  \citenamefont {Whitfield}, \citenamefont {Boixo}, \citenamefont {Tempel},\
  and\ \citenamefont {Aspuru-Guzik}}]{Yung2014}%
  \BibitemOpen
  \bibfield  {author} {\bibinfo {author} {\bibfnamefont {M.-H.}\ \bibnamefont
  {Yung}}, \bibinfo {author} {\bibfnamefont {J.~D.}\ \bibnamefont {Whitfield}},
  \bibinfo {author} {\bibfnamefont {S.}~\bibnamefont {Boixo}}, \bibinfo
  {author} {\bibfnamefont {D.~G.}\ \bibnamefont {Tempel}},\ and\ \bibinfo
  {author} {\bibfnamefont {A.}~\bibnamefont {Aspuru-Guzik}},\ }\bibinfo {title}
  {{Introduction to Quantum Algorithms for Physics and Chemistry}},\ in\ \href
  {https://doi.org/https://doi.org/10.1002/9781118742631.ch03} {\emph {\bibinfo
  {booktitle} {Quantum Information and Computation for Chemistry}}}\ (\bibinfo
  {publisher} {John Wiley \& Sons, Ltd},\ \bibinfo {year} {2014})\
  Chap.~\bibinfo {chapter} {3}, pp.\ \bibinfo {pages} {67--106}\BibitemShut
  {NoStop}%
\bibitem [{\citenamefont {Kitaev}(1996)}]{Kitaev1996}%
  \BibitemOpen
  \bibfield  {author} {\bibinfo {author} {\bibfnamefont {A.}~\bibnamefont
  {Kitaev}},\ }\bibfield  {title} {\bibinfo {title} {Quantum measurements and
  the {A}belian stabilizer problem},\ }\href@noop {} {\bibfield  {journal}
  {\bibinfo  {journal} {Electr. Coll. Comput. Complex.}\ }\textbf {\bibinfo
  {volume} {{TR96-003}}} (\bibinfo {year} {1996})}\BibitemShut {NoStop}%
\bibitem [{\citenamefont {Somma}\ \emph {et~al.}(2002)\citenamefont {Somma},
  \citenamefont {Ortiz}, \citenamefont {Gubernatis}, \citenamefont {Knill},\
  and\ \citenamefont {Laflamme}}]{Somma2002}%
  \BibitemOpen
  \bibfield  {author} {\bibinfo {author} {\bibfnamefont {R.}~\bibnamefont
  {Somma}}, \bibinfo {author} {\bibfnamefont {G.}~\bibnamefont {Ortiz}},
  \bibinfo {author} {\bibfnamefont {J.~E.}\ \bibnamefont {Gubernatis}},
  \bibinfo {author} {\bibfnamefont {E.}~\bibnamefont {Knill}},\ and\ \bibinfo
  {author} {\bibfnamefont {R.}~\bibnamefont {Laflamme}},\ }\bibfield  {title}
  {\bibinfo {title} {Simulating physical phenomena by quantum networks},\
  }\href {https://doi.org/10.1103/PhysRevA.65.042323} {\bibfield  {journal}
  {\bibinfo  {journal} {Phys. Rev. A}\ }\textbf {\bibinfo {volume} {65}},\
  \bibinfo {pages} {042323} (\bibinfo {year} {2002})}\BibitemShut {NoStop}%
\bibitem [{\citenamefont {Somma}(2019)}]{Somma2019}%
  \BibitemOpen
  \bibfield  {author} {\bibinfo {author} {\bibfnamefont {R.~D.}\ \bibnamefont
  {Somma}},\ }\bibfield  {title} {\bibinfo {title} {Quantum eigenvalue
  estimation via time series analysis},\ }\href
  {https://doi.org/10.1088/1367-2630/ab5c60} {\bibfield  {journal} {\bibinfo
  {journal} {New J. Phys.}\ }\textbf {\bibinfo {volume} {21}},\ \bibinfo
  {pages} {123025} (\bibinfo {year} {2019})}\BibitemShut {NoStop}%
\bibitem [{\citenamefont {O’Brien}\ \emph {et~al.}(2019)\citenamefont
  {O’Brien}, \citenamefont {Tarasinski},\ and\ \citenamefont
  {Terhal}}]{OBrien2019a}%
  \BibitemOpen
  \bibfield  {author} {\bibinfo {author} {\bibfnamefont {T.~E.}\ \bibnamefont
  {O’Brien}}, \bibinfo {author} {\bibfnamefont {B.}~\bibnamefont
  {Tarasinski}},\ and\ \bibinfo {author} {\bibfnamefont {B.~M.}\ \bibnamefont
  {Terhal}},\ }\bibfield  {title} {\bibinfo {title} {Quantum phase estimation
  of multiple eigenvalues for small-scale (noisy) experiments},\ }\href
  {https://doi.org/10.1088/1367-2630/aafb8e} {\bibfield  {journal} {\bibinfo
  {journal} {New J. Phys.}\ }\textbf {\bibinfo {volume} {21}},\ \bibinfo
  {pages} {023022} (\bibinfo {year} {2019})}\BibitemShut {NoStop}%
\bibitem [{\citenamefont {O'Brien}\ \emph {et~al.}(2021)\citenamefont
  {O'Brien}, \citenamefont {Polla}, \citenamefont {Rubin}, \citenamefont
  {Huggins}, \citenamefont {McArdle}, \citenamefont {Boixo}, \citenamefont
  {McClean},\ and\ \citenamefont {Babbush}}]{Obrien2021}%
  \BibitemOpen
  \bibfield  {author} {\bibinfo {author} {\bibfnamefont {T.~E.}\ \bibnamefont
  {O'Brien}}, \bibinfo {author} {\bibfnamefont {S.}~\bibnamefont {Polla}},
  \bibinfo {author} {\bibfnamefont {N.~C.}\ \bibnamefont {Rubin}}, \bibinfo
  {author} {\bibfnamefont {W.~J.}\ \bibnamefont {Huggins}}, \bibinfo {author}
  {\bibfnamefont {S.}~\bibnamefont {McArdle}}, \bibinfo {author} {\bibfnamefont
  {S.}~\bibnamefont {Boixo}}, \bibinfo {author} {\bibfnamefont {J.~R.}\
  \bibnamefont {McClean}},\ and\ \bibinfo {author} {\bibfnamefont
  {R.}~\bibnamefont {Babbush}},\ }\bibfield  {title} {\bibinfo {title} {{Error
  Mitigation via Verified Phase Estimation}},\ }\href
  {https://doi.org/10.1103/PRXQuantum.2.020317} {\bibfield  {journal} {\bibinfo
   {journal} {PRX Quantum}\ }\textbf {\bibinfo {volume} {2}},\ \bibinfo {pages}
  {020317} (\bibinfo {year} {2021})}\BibitemShut {NoStop}%
\bibitem [{\citenamefont {Lin}\ and\ \citenamefont {Tong}(2022)}]{Lin2022}%
  \BibitemOpen
  \bibfield  {author} {\bibinfo {author} {\bibfnamefont {L.}~\bibnamefont
  {Lin}}\ and\ \bibinfo {author} {\bibfnamefont {Y.}~\bibnamefont {Tong}},\
  }\bibfield  {title} {\bibinfo {title} {{Heisenberg-Limited Ground-State
  Energy Estimation for Early Fault-Tolerant Quantum Computers}},\ }\href
  {https://doi.org/10.1103/PRXQuantum.3.010318} {\bibfield  {journal} {\bibinfo
   {journal} {PRX Quantum}\ }\textbf {\bibinfo {volume} {3}},\ \bibinfo {pages}
  {010318} (\bibinfo {year} {2022})}\BibitemShut {NoStop}%
\bibitem [{\citenamefont {Wang}\ \emph {et~al.}(2023)\citenamefont {Wang},
  \citenamefont {Fran{\c{c}}a}, \citenamefont {Zhang}, \citenamefont {Zhu},\
  and\ \citenamefont {Johnson}}]{Wang2023}%
  \BibitemOpen
  \bibfield  {author} {\bibinfo {author} {\bibfnamefont {G.}~\bibnamefont
  {Wang}}, \bibinfo {author} {\bibfnamefont {D.~S.}\ \bibnamefont
  {Fran{\c{c}}a}}, \bibinfo {author} {\bibfnamefont {R.}~\bibnamefont {Zhang}},
  \bibinfo {author} {\bibfnamefont {S.}~\bibnamefont {Zhu}},\ and\ \bibinfo
  {author} {\bibfnamefont {P.~D.}\ \bibnamefont {Johnson}},\ }\bibfield
  {title} {\bibinfo {title} {Quantum algorithm for ground state energy
  estimation using circuit depth with exponentially improved dependence on
  precision},\ }\href {https://doi.org/10.22331/q-2023-11-06-1167} {\bibfield
  {journal} {\bibinfo  {journal} {{Quantum}}\ }\textbf {\bibinfo {volume}
  {7}},\ \bibinfo {pages} {1167} (\bibinfo {year} {2023})}\BibitemShut
  {NoStop}%
\bibitem [{\citenamefont {Ding}\ and\ \citenamefont
  {Lin}(2023{\natexlab{a}})}]{Ding2023a}%
  \BibitemOpen
  \bibfield  {author} {\bibinfo {author} {\bibfnamefont {Z.}~\bibnamefont
  {Ding}}\ and\ \bibinfo {author} {\bibfnamefont {L.}~\bibnamefont {Lin}},\
  }\bibfield  {title} {\bibinfo {title} {Even shorter quantum circuit for phase
  estimation on early fault-tolerant quantum computers with applications to
  ground-state energy estimation},\ }\href
  {https://doi.org/10.1103/PRXQuantum.4.020331} {\bibfield  {journal} {\bibinfo
   {journal} {PRX Quantum}\ }\textbf {\bibinfo {volume} {4}},\ \bibinfo {pages}
  {020331} (\bibinfo {year} {2023}{\natexlab{a}})}\BibitemShut {NoStop}%
\bibitem [{\citenamefont {Ding}\ and\ \citenamefont
  {Lin}(2023{\natexlab{b}})}]{Ding2023b}%
  \BibitemOpen
  \bibfield  {author} {\bibinfo {author} {\bibfnamefont {Z.}~\bibnamefont
  {Ding}}\ and\ \bibinfo {author} {\bibfnamefont {L.}~\bibnamefont {Lin}},\
  }\bibfield  {title} {\bibinfo {title} {Simultaneous estimation of multiple
  eigenvalues with short-depth quantum circuit on early fault-tolerant quantum
  computers},\ }\href {https://doi.org/10.22331/q-2023-10-11-1136} {\bibfield
  {journal} {\bibinfo  {journal} {{Quantum}}\ }\textbf {\bibinfo {volume}
  {7}},\ \bibinfo {pages} {1136} (\bibinfo {year}
  {2023}{\natexlab{b}})}\BibitemShut {NoStop}%
\bibitem [{\citenamefont {Wecker}\ \emph
  {et~al.}(2015{\natexlab{a}})\citenamefont {Wecker}, \citenamefont {Hastings},
  \citenamefont {Wiebe}, \citenamefont {Clark}, \citenamefont {Nayak},\ and\
  \citenamefont {Troyer}}]{Wecker2015}%
  \BibitemOpen
  \bibfield  {author} {\bibinfo {author} {\bibfnamefont {D.}~\bibnamefont
  {Wecker}}, \bibinfo {author} {\bibfnamefont {M.~B.}\ \bibnamefont
  {Hastings}}, \bibinfo {author} {\bibfnamefont {N.}~\bibnamefont {Wiebe}},
  \bibinfo {author} {\bibfnamefont {B.~K.}\ \bibnamefont {Clark}}, \bibinfo
  {author} {\bibfnamefont {C.}~\bibnamefont {Nayak}},\ and\ \bibinfo {author}
  {\bibfnamefont {M.}~\bibnamefont {Troyer}},\ }\bibfield  {title} {\bibinfo
  {title} {Solving strongly correlated electron models on a quantum computer},\
  }\href {https://doi.org/10.1103/PhysRevA.92.062318} {\bibfield  {journal}
  {\bibinfo  {journal} {Phys. Rev. A}\ }\textbf {\bibinfo {volume} {92}},\
  \bibinfo {pages} {062318} (\bibinfo {year} {2015}{\natexlab{a}})}\BibitemShut
  {NoStop}%
\bibitem [{\citenamefont {Zintchenko}\ and\ \citenamefont
  {Wiebe}(2016)}]{Zintchenko2016}%
  \BibitemOpen
  \bibfield  {author} {\bibinfo {author} {\bibfnamefont {I.}~\bibnamefont
  {Zintchenko}}\ and\ \bibinfo {author} {\bibfnamefont {N.}~\bibnamefont
  {Wiebe}},\ }\bibfield  {title} {\bibinfo {title} {Randomized gap and
  amplitude estimation},\ }\href {https://doi.org/10.1103/PhysRevA.93.062306}
  {\bibfield  {journal} {\bibinfo  {journal} {Phys. Rev. A}\ }\textbf {\bibinfo
  {volume} {93}},\ \bibinfo {pages} {062306} (\bibinfo {year}
  {2016})}\BibitemShut {NoStop}%
\bibitem [{\citenamefont {Sugisaki}\ \emph {et~al.}(2021)\citenamefont
  {Sugisaki}, \citenamefont {Sakai}, \citenamefont {Toyota}, \citenamefont
  {Sato}, \citenamefont {Shiomi},\ and\ \citenamefont {Takui}}]{Sugisaki2021}%
  \BibitemOpen
  \bibfield  {author} {\bibinfo {author} {\bibfnamefont {K.}~\bibnamefont
  {Sugisaki}}, \bibinfo {author} {\bibfnamefont {C.}~\bibnamefont {Sakai}},
  \bibinfo {author} {\bibfnamefont {K.}~\bibnamefont {Toyota}}, \bibinfo
  {author} {\bibfnamefont {K.}~\bibnamefont {Sato}}, \bibinfo {author}
  {\bibfnamefont {D.}~\bibnamefont {Shiomi}},\ and\ \bibinfo {author}
  {\bibfnamefont {T.}~\bibnamefont {Takui}},\ }\bibfield  {title} {\bibinfo
  {title} {Bayesian phase difference estimation: a general quantum algorithm
  for the direct calculation of energy gaps},\ }\href
  {https://doi.org/10.1039/D1CP03156B} {\bibfield  {journal} {\bibinfo
  {journal} {Phys. Chem. Chem. Phys.}\ }\textbf {\bibinfo {volume} {23}},\
  \bibinfo {pages} {20152} (\bibinfo {year} {2021})}\BibitemShut {NoStop}%
\bibitem [{\citenamefont {Russo}\ \emph {et~al.}(2021)\citenamefont {Russo},
  \citenamefont {Rudinger}, \citenamefont {Morrison},\ and\ \citenamefont
  {Baczewski}}]{Russo2021}%
  \BibitemOpen
  \bibfield  {author} {\bibinfo {author} {\bibfnamefont {A.~E.}\ \bibnamefont
  {Russo}}, \bibinfo {author} {\bibfnamefont {K.~M.}\ \bibnamefont {Rudinger}},
  \bibinfo {author} {\bibfnamefont {B.~C.~A.}\ \bibnamefont {Morrison}},\ and\
  \bibinfo {author} {\bibfnamefont {A.~D.}\ \bibnamefont {Baczewski}},\
  }\bibfield  {title} {\bibinfo {title} {{Evaluating Energy Differences on a
  Quantum Computer with Robust Phase Estimation}},\ }\href
  {https://doi.org/10.1103/PhysRevLett.126.210501} {\bibfield  {journal}
  {\bibinfo  {journal} {Phys. Rev. Lett.}\ }\textbf {\bibinfo {volume} {126}},\
  \bibinfo {pages} {210501} (\bibinfo {year} {2021})}\BibitemShut {NoStop}%
\bibitem [{\citenamefont {Matsuzaki}\ \emph {et~al.}(2021)\citenamefont
  {Matsuzaki}, \citenamefont {Hakoshima}, \citenamefont {Sugisaki},
  \citenamefont {Seki},\ and\ \citenamefont {Kawabata}}]{Matsuzaki2021}%
  \BibitemOpen
  \bibfield  {author} {\bibinfo {author} {\bibfnamefont {Y.}~\bibnamefont
  {Matsuzaki}}, \bibinfo {author} {\bibfnamefont {H.}~\bibnamefont
  {Hakoshima}}, \bibinfo {author} {\bibfnamefont {K.}~\bibnamefont {Sugisaki}},
  \bibinfo {author} {\bibfnamefont {Y.}~\bibnamefont {Seki}},\ and\ \bibinfo
  {author} {\bibfnamefont {S.}~\bibnamefont {Kawabata}},\ }\bibfield  {title}
  {\bibinfo {title} {Direct estimation of the energy gap between the ground
  state and excited state with quantum annealing},\ }\href
  {https://doi.org/10.35848/1347-4065/abdf20} {\bibfield  {journal} {\bibinfo
  {journal} {Jpn. J. Appl. Phys.}\ }\textbf {\bibinfo {volume} {60}},\ \bibinfo
  {pages} {SBBI02} (\bibinfo {year} {2021})}\BibitemShut {NoStop}%
\bibitem [{\citenamefont {Trotter}(1959)}]{Trotter1959}%
  \BibitemOpen
  \bibfield  {author} {\bibinfo {author} {\bibfnamefont {H.~F.}\ \bibnamefont
  {Trotter}},\ }\bibfield  {title} {\bibinfo {title} {On the product of
  semi-groups of operators},\ }\href {http://www.jstor.org/stable/2033649}
  {\bibfield  {journal} {\bibinfo  {journal} {Proc. Am. Math. Soc.}\ }\textbf
  {\bibinfo {volume} {10}},\ \bibinfo {pages} {545} (\bibinfo {year}
  {1959})}\BibitemShut {NoStop}%
\bibitem [{\citenamefont {Suzuki}(1976)}]{Suzuki1976}%
  \BibitemOpen
  \bibfield  {author} {\bibinfo {author} {\bibfnamefont {M.}~\bibnamefont
  {Suzuki}},\ }\bibfield  {title} {\bibinfo {title} {Generalized {T}rotter's
  formula and systematic approximants of exponential operators and inner
  derivations with applications to many-body problems},\ }\href
  {https://doi.org/10.1007/BF01609348} {\bibfield  {journal} {\bibinfo
  {journal} {Commun. Math. Phys.}\ }\textbf {\bibinfo {volume} {51}},\ \bibinfo
  {pages} {183} (\bibinfo {year} {1976})}\BibitemShut {NoStop}%
\bibitem [{\citenamefont {Barthel}\ and\ \citenamefont
  {Zhang}(2020)}]{Barthel2020}%
  \BibitemOpen
  \bibfield  {author} {\bibinfo {author} {\bibfnamefont {T.}~\bibnamefont
  {Barthel}}\ and\ \bibinfo {author} {\bibfnamefont {Y.}~\bibnamefont
  {Zhang}},\ }\bibfield  {title} {\bibinfo {title} {Optimized
  {Lie–Trotter–Suzuki} decompositions for two and three non-commuting
  terms},\ }\href {https://doi.org/https://doi.org/10.1016/j.aop.2020.168165}
  {\bibfield  {journal} {\bibinfo  {journal} {Ann. Phys.}\ }\textbf {\bibinfo
  {volume} {418}},\ \bibinfo {pages} {168165} (\bibinfo {year}
  {2020})}\BibitemShut {NoStop}%
\bibitem [{\citenamefont {Childs}\ \emph {et~al.}(2021)\citenamefont {Childs},
  \citenamefont {Su}, \citenamefont {Tran}, \citenamefont {Wiebe},\ and\
  \citenamefont {Zhu}}]{Childs2021}%
  \BibitemOpen
  \bibfield  {author} {\bibinfo {author} {\bibfnamefont {A.~M.}\ \bibnamefont
  {Childs}}, \bibinfo {author} {\bibfnamefont {Y.}~\bibnamefont {Su}}, \bibinfo
  {author} {\bibfnamefont {M.~C.}\ \bibnamefont {Tran}}, \bibinfo {author}
  {\bibfnamefont {N.}~\bibnamefont {Wiebe}},\ and\ \bibinfo {author}
  {\bibfnamefont {S.}~\bibnamefont {Zhu}},\ }\bibfield  {title} {\bibinfo
  {title} {{Theory of {Trotter} Error with Commutator Scaling}},\ }\href
  {https://doi.org/10.1103/PhysRevX.11.011020} {\bibfield  {journal} {\bibinfo
  {journal} {Phys. Rev. X}\ }\textbf {\bibinfo {volume} {11}},\ \bibinfo
  {pages} {011020} (\bibinfo {year} {2021})}\BibitemShut {NoStop}%
\bibitem [{\citenamefont {Peruzzo}\ \emph {et~al.}(2014)\citenamefont
  {Peruzzo}, \citenamefont {McClean}, \citenamefont {Shadbolt}, \citenamefont
  {Yung}, \citenamefont {Zhou}, \citenamefont {Love}, \citenamefont
  {Aspuru-Guzik},\ and\ \citenamefont {O'Brien}}]{Peruzzo2014}%
  \BibitemOpen
  \bibfield  {author} {\bibinfo {author} {\bibfnamefont {A.}~\bibnamefont
  {Peruzzo}}, \bibinfo {author} {\bibfnamefont {J.}~\bibnamefont {McClean}},
  \bibinfo {author} {\bibfnamefont {P.}~\bibnamefont {Shadbolt}}, \bibinfo
  {author} {\bibfnamefont {M.-H.}\ \bibnamefont {Yung}}, \bibinfo {author}
  {\bibfnamefont {X.-Q.}\ \bibnamefont {Zhou}}, \bibinfo {author}
  {\bibfnamefont {P.~J.}\ \bibnamefont {Love}}, \bibinfo {author}
  {\bibfnamefont {A.}~\bibnamefont {Aspuru-Guzik}},\ and\ \bibinfo {author}
  {\bibfnamefont {J.~L.}\ \bibnamefont {O'Brien}},\ }\bibfield  {title}
  {\bibinfo {title} {A variational eigenvalue solver on a photonic quantum
  processor},\ }\href {https://doi.org/10.1038/ncomms5213} {\bibfield
  {journal} {\bibinfo  {journal} {Nat. Commun.}\ }\textbf {\bibinfo {volume}
  {5}},\ \bibinfo {pages} {4213} (\bibinfo {year} {2014})}\BibitemShut
  {NoStop}%
\bibitem [{\citenamefont {McClean}\ \emph {et~al.}(2014)\citenamefont
  {McClean}, \citenamefont {Babbush}, \citenamefont {Love},\ and\ \citenamefont
  {Aspuru-Guzik}}]{McClean2014}%
  \BibitemOpen
  \bibfield  {author} {\bibinfo {author} {\bibfnamefont {J.~R.}\ \bibnamefont
  {McClean}}, \bibinfo {author} {\bibfnamefont {R.}~\bibnamefont {Babbush}},
  \bibinfo {author} {\bibfnamefont {P.~J.}\ \bibnamefont {Love}},\ and\
  \bibinfo {author} {\bibfnamefont {A.}~\bibnamefont {Aspuru-Guzik}},\
  }\bibfield  {title} {\bibinfo {title} {Exploiting locality in quantum
  computation for quantum chemistry},\ }\href
  {https://doi.org/10.1021/jz501649m} {\bibfield  {journal} {\bibinfo
  {journal} {J. Phys. Chem. Lett.}\ }\textbf {\bibinfo {volume} {5}},\ \bibinfo
  {pages} {4368} (\bibinfo {year} {2014})}\BibitemShut {NoStop}%
\bibitem [{\citenamefont {Wecker}\ \emph
  {et~al.}(2015{\natexlab{b}})\citenamefont {Wecker}, \citenamefont
  {Hastings},\ and\ \citenamefont {Troyer}}]{Wecker2015a}%
  \BibitemOpen
  \bibfield  {author} {\bibinfo {author} {\bibfnamefont {D.}~\bibnamefont
  {Wecker}}, \bibinfo {author} {\bibfnamefont {M.~B.}\ \bibnamefont
  {Hastings}},\ and\ \bibinfo {author} {\bibfnamefont {M.}~\bibnamefont
  {Troyer}},\ }\bibfield  {title} {\bibinfo {title} {Progress towards practical
  quantum variational algorithms},\ }\href
  {https://doi.org/10.1103/PhysRevA.92.042303} {\bibfield  {journal} {\bibinfo
  {journal} {Phys. Rev. A}\ }\textbf {\bibinfo {volume} {92}},\ \bibinfo
  {pages} {042303} (\bibinfo {year} {2015}{\natexlab{b}})}\BibitemShut
  {NoStop}%
\bibitem [{\citenamefont {McClean}\ \emph {et~al.}(2016)\citenamefont
  {McClean}, \citenamefont {Romero}, \citenamefont {Babbush},\ and\
  \citenamefont {Aspuru-Guzik}}]{McClean2016}%
  \BibitemOpen
  \bibfield  {author} {\bibinfo {author} {\bibfnamefont {J.~R.}\ \bibnamefont
  {McClean}}, \bibinfo {author} {\bibfnamefont {J.}~\bibnamefont {Romero}},
  \bibinfo {author} {\bibfnamefont {R.}~\bibnamefont {Babbush}},\ and\ \bibinfo
  {author} {\bibfnamefont {A.}~\bibnamefont {Aspuru-Guzik}},\ }\bibfield
  {title} {\bibinfo {title} {The theory of variational hybrid quantum-classical
  algorithms},\ }\href {https://doi.org/10.1088/1367-2630/18/2/023023}
  {\bibfield  {journal} {\bibinfo  {journal} {New J. Phys.}\ }\textbf {\bibinfo
  {volume} {18}},\ \bibinfo {pages} {023023} (\bibinfo {year}
  {2016})}\BibitemShut {NoStop}%
\bibitem [{\citenamefont {Tilly}\ \emph {et~al.}(2022)\citenamefont {Tilly},
  \citenamefont {Chen}, \citenamefont {Cao}, \citenamefont {Picozzi},
  \citenamefont {Setia}, \citenamefont {Li}, \citenamefont {Grant},
  \citenamefont {Wossnig}, \citenamefont {Rungger}, \citenamefont {Booth},\
  and\ \citenamefont {Tennyson}}]{Tilly2022}%
  \BibitemOpen
  \bibfield  {author} {\bibinfo {author} {\bibfnamefont {J.}~\bibnamefont
  {Tilly}}, \bibinfo {author} {\bibfnamefont {H.}~\bibnamefont {Chen}},
  \bibinfo {author} {\bibfnamefont {S.}~\bibnamefont {Cao}}, \bibinfo {author}
  {\bibfnamefont {D.}~\bibnamefont {Picozzi}}, \bibinfo {author} {\bibfnamefont
  {K.}~\bibnamefont {Setia}}, \bibinfo {author} {\bibfnamefont
  {Y.}~\bibnamefont {Li}}, \bibinfo {author} {\bibfnamefont {E.}~\bibnamefont
  {Grant}}, \bibinfo {author} {\bibfnamefont {L.}~\bibnamefont {Wossnig}},
  \bibinfo {author} {\bibfnamefont {I.}~\bibnamefont {Rungger}}, \bibinfo
  {author} {\bibfnamefont {G.~H.}\ \bibnamefont {Booth}},\ and\ \bibinfo
  {author} {\bibfnamefont {J.}~\bibnamefont {Tennyson}},\ }\bibfield  {title}
  {\bibinfo {title} {The variational quantum eigensolver: A review of methods
  and best practices},\ }\href
  {https://doi.org/https://doi.org/10.1016/j.physrep.2022.08.003} {\bibfield
  {journal} {\bibinfo  {journal} {Phys. Rep.}\ }\textbf {\bibinfo {volume}
  {986}},\ \bibinfo {pages} {1} (\bibinfo {year} {2022})}\BibitemShut {NoStop}%
\bibitem [{\citenamefont {McClean}\ \emph {et~al.}(2018)\citenamefont
  {McClean}, \citenamefont {Boixo}, \citenamefont {Smelyanskiy}, \citenamefont
  {Babbush},\ and\ \citenamefont {Neven}}]{McClean2018}%
  \BibitemOpen
  \bibfield  {author} {\bibinfo {author} {\bibfnamefont {J.~R.}\ \bibnamefont
  {McClean}}, \bibinfo {author} {\bibfnamefont {S.}~\bibnamefont {Boixo}},
  \bibinfo {author} {\bibfnamefont {V.~N.}\ \bibnamefont {Smelyanskiy}},
  \bibinfo {author} {\bibfnamefont {R.}~\bibnamefont {Babbush}},\ and\ \bibinfo
  {author} {\bibfnamefont {H.}~\bibnamefont {Neven}},\ }\bibfield  {title}
  {\bibinfo {title} {Barren plateaus in quantum neural network training
  landscapes},\ }\href {https://doi.org/10.1038/s41467-018-07090-4} {\bibfield
  {journal} {\bibinfo  {journal} {Nat. Commun.}\ }\textbf {\bibinfo {volume}
  {9}},\ \bibinfo {pages} {4812} (\bibinfo {year} {2018})}\BibitemShut
  {NoStop}%
\bibitem [{\citenamefont {Lee}\ \emph {et~al.}(2021)\citenamefont {Lee},
  \citenamefont {Berry}, \citenamefont {Gidney}, \citenamefont {Huggins},
  \citenamefont {McClean}, \citenamefont {Wiebe},\ and\ \citenamefont
  {Babbush}}]{Lee2021}%
  \BibitemOpen
  \bibfield  {author} {\bibinfo {author} {\bibfnamefont {J.}~\bibnamefont
  {Lee}}, \bibinfo {author} {\bibfnamefont {D.~W.}\ \bibnamefont {Berry}},
  \bibinfo {author} {\bibfnamefont {C.}~\bibnamefont {Gidney}}, \bibinfo
  {author} {\bibfnamefont {W.~J.}\ \bibnamefont {Huggins}}, \bibinfo {author}
  {\bibfnamefont {J.~R.}\ \bibnamefont {McClean}}, \bibinfo {author}
  {\bibfnamefont {N.}~\bibnamefont {Wiebe}},\ and\ \bibinfo {author}
  {\bibfnamefont {R.}~\bibnamefont {Babbush}},\ }\bibfield  {title} {\bibinfo
  {title} {{Even More Efficient Quantum Computations of Chemistry Through
  Tensor Hypercontraction}},\ }\href
  {https://doi.org/10.1103/PRXQuantum.2.030305} {\bibfield  {journal} {\bibinfo
   {journal} {PRX Quantum}\ }\textbf {\bibinfo {volume} {2}},\ \bibinfo {pages}
  {030305} (\bibinfo {year} {2021})}\BibitemShut {NoStop}%
\bibitem [{\citenamefont {Terhal}(2015)}]{Tehral2015}%
  \BibitemOpen
  \bibfield  {author} {\bibinfo {author} {\bibfnamefont {B.~M.}\ \bibnamefont
  {Terhal}},\ }\bibfield  {title} {\bibinfo {title} {Quantum error correction
  for quantum memories},\ }\href {https://doi.org/10.1103/RevModPhys.87.307}
  {\bibfield  {journal} {\bibinfo  {journal} {Rev. Mod. Phys.}\ }\textbf
  {\bibinfo {volume} {87}},\ \bibinfo {pages} {307} (\bibinfo {year}
  {2015})}\BibitemShut {NoStop}%
\bibitem [{\citenamefont {Bauer}\ \emph {et~al.}(2016)\citenamefont {Bauer},
  \citenamefont {Wecker}, \citenamefont {Millis}, \citenamefont {Hastings},\
  and\ \citenamefont {Troyer}}]{Bauer2016}%
  \BibitemOpen
  \bibfield  {author} {\bibinfo {author} {\bibfnamefont {B.}~\bibnamefont
  {Bauer}}, \bibinfo {author} {\bibfnamefont {D.}~\bibnamefont {Wecker}},
  \bibinfo {author} {\bibfnamefont {A.~J.}\ \bibnamefont {Millis}}, \bibinfo
  {author} {\bibfnamefont {M.~B.}\ \bibnamefont {Hastings}},\ and\ \bibinfo
  {author} {\bibfnamefont {M.}~\bibnamefont {Troyer}},\ }\bibfield  {title}
  {\bibinfo {title} {{Hybrid Quantum-Classical Approach to Correlated
  Materials}},\ }\href {https://doi.org/10.1103/PhysRevX.6.031045} {\bibfield
  {journal} {\bibinfo  {journal} {Phys. Rev. X}\ }\textbf {\bibinfo {volume}
  {6}},\ \bibinfo {pages} {031045} (\bibinfo {year} {2016})}\BibitemShut
  {NoStop}%
\bibitem [{\citenamefont {Keen}\ \emph {et~al.}(2020)\citenamefont {Keen},
  \citenamefont {Maier}, \citenamefont {Johnston},\ and\ \citenamefont
  {Lougovski}}]{Keen2020}%
  \BibitemOpen
  \bibfield  {author} {\bibinfo {author} {\bibfnamefont {T.}~\bibnamefont
  {Keen}}, \bibinfo {author} {\bibfnamefont {T.}~\bibnamefont {Maier}},
  \bibinfo {author} {\bibfnamefont {S.}~\bibnamefont {Johnston}},\ and\
  \bibinfo {author} {\bibfnamefont {P.}~\bibnamefont {Lougovski}},\ }\bibfield
  {title} {\bibinfo {title} {Quantum-classical simulation of two-site dynamical
  mean-field theory on noisy quantum hardware},\ }\href
  {https://doi.org/10.1088/2058-9565/ab7d4c} {\bibfield  {journal} {\bibinfo
  {journal} {Quantum Sci. Technol.}\ }\textbf {\bibinfo {volume} {5}},\
  \bibinfo {pages} {035001} (\bibinfo {year} {2020})}\BibitemShut {NoStop}%
\bibitem [{\citenamefont {Steckmann}\ \emph {et~al.}(2023)\citenamefont
  {Steckmann}, \citenamefont {Keen}, \citenamefont {Kökcü}, \citenamefont
  {Kemper}, \citenamefont {Dumitrescu},\ and\ \citenamefont
  {Wang}}]{Steckmann2023}%
  \BibitemOpen
  \bibfield  {author} {\bibinfo {author} {\bibfnamefont {T.}~\bibnamefont
  {Steckmann}}, \bibinfo {author} {\bibfnamefont {T.}~\bibnamefont {Keen}},
  \bibinfo {author} {\bibfnamefont {E.}~\bibnamefont {Kökcü}}, \bibinfo
  {author} {\bibfnamefont {A.~F.}\ \bibnamefont {Kemper}}, \bibinfo {author}
  {\bibfnamefont {E.~F.}\ \bibnamefont {Dumitrescu}},\ and\ \bibinfo {author}
  {\bibfnamefont {Y.}~\bibnamefont {Wang}},\ }\bibfield  {title} {\bibinfo
  {title} {Mapping the metal-insulator phase diagram by algebraically
  fast-forwarding dynamics on a cloud quantum computer},\ }\href@noop {}
  {\bibfield  {journal} {\bibinfo  {journal} {arXiv:2112.05688 [quant-ph]}\ }
  (\bibinfo {year} {2023})}\BibitemShut {NoStop}%
\bibitem [{\citenamefont {Lee}\ \emph {et~al.}()\citenamefont {Lee},
  \citenamefont {Myers},\ and\ \citenamefont {Scarola}}]{Lee2024}%
  \BibitemOpen
  \bibfield  {author} {\bibinfo {author} {\bibfnamefont {W.-R.}\ \bibnamefont
  {Lee}}, \bibinfo {author} {\bibfnamefont {N.~M.}\ \bibnamefont {Myers}},\
  and\ \bibinfo {author} {\bibfnamefont {V.~W.}\ \bibnamefont {Scarola}},\
  }\bibfield  {title} {\bibinfo {title} {Self-embedded fault tolerance in
  quantum gap estimation with trial-state optimization},\ }\href@noop {}
  {\bibinfo  {journal} {unpublished}\ }\BibitemShut {NoStop}%
\bibitem [{\citenamefont {Childs}\ \emph {et~al.}(2018)\citenamefont {Childs},
  \citenamefont {Maslov}, \citenamefont {Nam}, \citenamefont {Ross},\ and\
  \citenamefont {Su}}]{Childs2018}%
  \BibitemOpen
\bibfield  {journal} {  }\bibfield  {author} {\bibinfo {author} {\bibfnamefont
  {A.~M.}\ \bibnamefont {Childs}}, \bibinfo {author} {\bibfnamefont
  {D.}~\bibnamefont {Maslov}}, \bibinfo {author} {\bibfnamefont
  {Y.}~\bibnamefont {Nam}}, \bibinfo {author} {\bibfnamefont {N.~J.}\
  \bibnamefont {Ross}},\ and\ \bibinfo {author} {\bibfnamefont
  {Y.}~\bibnamefont {Su}},\ }\bibfield  {title} {\bibinfo {title} {Toward the
  first quantum simulation with quantum speedup},\ }\href
  {https://doi.org/10.1073/pnas.1801723115} {\bibfield  {journal} {\bibinfo
  {journal} {Proc. Natl. Acad. Sci. U.S.A.}\ }\textbf {\bibinfo {volume}
  {115}},\ \bibinfo {pages} {9456} (\bibinfo {year} {2018})}\BibitemShut
  {NoStop}%
\bibitem [{\citenamefont {Childs}\ and\ \citenamefont {Su}(2019)}]{Childs2019}%
  \BibitemOpen
  \bibfield  {author} {\bibinfo {author} {\bibfnamefont {A.~M.}\ \bibnamefont
  {Childs}}\ and\ \bibinfo {author} {\bibfnamefont {Y.}~\bibnamefont {Su}},\
  }\bibfield  {title} {\bibinfo {title} {{Nearly Optimal Lattice Simulation by
  Product Formulas}},\ }\href {https://doi.org/10.1103/PhysRevLett.123.050503}
  {\bibfield  {journal} {\bibinfo  {journal} {Phys. Rev. Lett.}\ }\textbf
  {\bibinfo {volume} {123}},\ \bibinfo {pages} {050503} (\bibinfo {year}
  {2019})}\BibitemShut {NoStop}%
\bibitem [{\citenamefont {Tran}\ \emph {et~al.}(2020)\citenamefont {Tran},
  \citenamefont {Chu}, \citenamefont {Su}, \citenamefont {Childs},\ and\
  \citenamefont {Gorshkov}}]{Tran2020}%
  \BibitemOpen
  \bibfield  {author} {\bibinfo {author} {\bibfnamefont {M.~C.}\ \bibnamefont
  {Tran}}, \bibinfo {author} {\bibfnamefont {S.~K.}\ \bibnamefont {Chu}},
  \bibinfo {author} {\bibfnamefont {Y.}~\bibnamefont {Su}}, \bibinfo {author}
  {\bibfnamefont {A.~M.}\ \bibnamefont {Childs}},\ and\ \bibinfo {author}
  {\bibfnamefont {A.~V.}\ \bibnamefont {Gorshkov}},\ }\bibfield  {title}
  {\bibinfo {title} {{Destructive Error Interference in Product-Formula Lattice
  Simulation}},\ }\href {https://doi.org/10.1103/PhysRevLett.124.220502}
  {\bibfield  {journal} {\bibinfo  {journal} {Phys. Rev. Lett.}\ }\textbf
  {\bibinfo {volume} {124}},\ \bibinfo {pages} {220502} (\bibinfo {year}
  {2020})}\BibitemShut {NoStop}%
\bibitem [{\citenamefont {Haug}\ and\ \citenamefont {Jauho}(2007)}]{Haug2007}%
  \BibitemOpen
  \bibfield  {author} {\bibinfo {author} {\bibfnamefont {H.}~\bibnamefont
  {Haug}}\ and\ \bibinfo {author} {\bibfnamefont {A.}~\bibnamefont {Jauho}},\
  }\href {https://books.google.com/books?id=w1am24ZE9jQC} {\emph {\bibinfo
  {title} {Quantum Kinetics in Transport and Optics of Semiconductors}}},\
  Springer Series in Solid-State Sciences\ (\bibinfo  {publisher} {Springer
  Berlin Heidelberg},\ \bibinfo {year} {2007})\BibitemShut {NoStop}%
\bibitem [{Com({\natexlab{a}})}]{Comment_selfenergy}%
  \BibitemOpen
  \bibinfo {note} {The self-energy may be set up to describe spin-dependent
  scattering processes. Here we avoid these cases since they generally need
  additional Trotterization for circuit implementation.}\BibitemShut {Stop}%
\bibitem [{\citenamefont {Sachdev}(2011)}]{Sachdev2011}%
  \BibitemOpen
  \bibfield  {author} {\bibinfo {author} {\bibfnamefont {S.}~\bibnamefont
  {Sachdev}},\ }\href {https://books.google.com/books?id=F3IkpxwpqSgC} {\emph
  {\bibinfo {title} {\textit{Quantum Phase Transitions}}}}\ (\bibinfo
  {publisher} {Cambridge University Press},\ \bibinfo {year}
  {2011})\BibitemShut {NoStop}%
\bibitem [{\citenamefont {Vogel}\ and\ \citenamefont
  {Risken}(1989)}]{Vogel1989}%
  \BibitemOpen
  \bibfield  {author} {\bibinfo {author} {\bibfnamefont {K.}~\bibnamefont
  {Vogel}}\ and\ \bibinfo {author} {\bibfnamefont {H.}~\bibnamefont {Risken}},\
  }\bibfield  {title} {\bibinfo {title} {Determination of quasiprobability
  distributions in terms of probability distributions for the rotated
  quadrature phase},\ }\href {https://doi.org/10.1103/PhysRevA.40.2847}
  {\bibfield  {journal} {\bibinfo  {journal} {Phys. Rev. A}\ }\textbf {\bibinfo
  {volume} {40}},\ \bibinfo {pages} {2847} (\bibinfo {year}
  {1989})}\BibitemShut {NoStop}%
\bibitem [{\citenamefont {Duhamel}\ and\ \citenamefont
  {Vetterli}(1990)}]{Duhamel1990}%
  \BibitemOpen
  \bibfield  {author} {\bibinfo {author} {\bibfnamefont {P.}~\bibnamefont
  {Duhamel}}\ and\ \bibinfo {author} {\bibfnamefont {M.}~\bibnamefont
  {Vetterli}},\ }\bibfield  {title} {\bibinfo {title} {Fast fourier transforms:
  A tutorial review and a state of the art},\ }\href
  {https://doi.org/https://doi.org/10.1016/0165-1684(90)90158-U} {\bibfield
  {journal} {\bibinfo  {journal} {Signal Process.}\ }\textbf {\bibinfo {volume}
  {19}},\ \bibinfo {pages} {259} (\bibinfo {year} {1990})}\BibitemShut
  {NoStop}%
\bibitem [{Com({\natexlab{b}})}]{Comment_QEE}%
  \BibitemOpen
  \bibinfo {note} {In contrast to QGE, the time propagator for eigenvalue
  estimation is not represented in a density matrix form. For eigenvalue
  estimation, instead, a Hadamard test circuit is constructed to return the
  real and imaginary parts of the time propagator when a single ancilla qubit
  is measured (See Refs.~\cite{Somma2002,Lin2022}, for example).}\BibitemShut
  {Stop}%
\bibitem [{Com({\natexlab{c}})}]{Comment_GitHub}%
  \BibitemOpen
  \bibinfo {note} {Qiskit code for the hybrid QGE algorithm is provided in: \\
  https://github.com/wrlee7609/hybrid\_quantum\_gap\_estimation.}\BibitemShut
  {Stop}%
\bibitem [{Com({\natexlab{d}})}]{Comment_parameters}%
  \BibitemOpen
  \bibinfo {note} {For the filter $\mathcal{F}$ with the broadening $\eta$, it
  is enough to choose the frequency unit $\delta\omega=\eta/4$ implying that
  only 4 sampling points are contained within the halfwidth of a peak. The
  sampling size is set to $L = 2\lceil 7h/\delta\omega\rceil$ yielding the
  frequency window larger than the bandwidth of the many-body model in our
  consideration. To get better precision in gap estimation, the sampling data
  can be interpolated whatever we take for ($\mathcal{F}$,
  $\eta$).}\BibitemShut {Stop}%
\bibitem [{\citenamefont {Heyl}\ \emph {et~al.}(2019)\citenamefont {Heyl},
  \citenamefont {Hauke},\ and\ \citenamefont {Zoller}}]{Heyl2019}%
  \BibitemOpen
  \bibfield  {author} {\bibinfo {author} {\bibfnamefont {M.}~\bibnamefont
  {Heyl}}, \bibinfo {author} {\bibfnamefont {P.}~\bibnamefont {Hauke}},\ and\
  \bibinfo {author} {\bibfnamefont {P.}~\bibnamefont {Zoller}},\ }\bibfield
  {title} {\bibinfo {title} {Quantum localization bounds {Trotter} errors in
  digital quantum simulation},\ }\href {https://doi.org/10.1126/sciadv.aau8342}
  {\bibfield  {journal} {\bibinfo  {journal} {Sci. Adv.}\ }\textbf {\bibinfo
  {volume} {5}},\ \bibinfo {pages} {eaau8342} (\bibinfo {year}
  {2019})}\BibitemShut {NoStop}%
\bibitem [{\citenamefont {Chicco}\ \emph {et~al.}(2021)\citenamefont {Chicco},
  \citenamefont {Warrens},\ and\ \citenamefont {Jurman}}]{Chicco2021}%
  \BibitemOpen
  \bibfield  {author} {\bibinfo {author} {\bibfnamefont {D.}~\bibnamefont
  {Chicco}}, \bibinfo {author} {\bibfnamefont {M.~J.}\ \bibnamefont
  {Warrens}},\ and\ \bibinfo {author} {\bibfnamefont {G.}~\bibnamefont
  {Jurman}},\ }\bibfield  {title} {\bibinfo {title} {{The coefficient of
  determination R-squared is more informative than SMAPE, MAE, MAPE, MSE and
  RMSE in regression analysis evaluation}},\ }\href
  {https://doi.org/10.7717/peerj-cs.623} {\bibfield  {journal} {\bibinfo
  {journal} {{PeerJ Comput. Sci.}}\ }\textbf {\bibinfo {volume} {7}},\ \bibinfo
  {pages} {e623} (\bibinfo {year} {2021})}\BibitemShut {NoStop}%
\bibitem [{\citenamefont {K\"okc\"u}\ \emph {et~al.}(2022)\citenamefont
  {K\"okc\"u}, \citenamefont {Steckmann}, \citenamefont {Wang}, \citenamefont
  {Freericks}, \citenamefont {Dumitrescu},\ and\ \citenamefont
  {Kemper}}]{Kokcu2022}%
  \BibitemOpen
  \bibfield  {author} {\bibinfo {author} {\bibfnamefont {E.}~\bibnamefont
  {K\"okc\"u}}, \bibinfo {author} {\bibfnamefont {T.}~\bibnamefont
  {Steckmann}}, \bibinfo {author} {\bibfnamefont {Y.}~\bibnamefont {Wang}},
  \bibinfo {author} {\bibfnamefont {J.~K.}\ \bibnamefont {Freericks}}, \bibinfo
  {author} {\bibfnamefont {E.~F.}\ \bibnamefont {Dumitrescu}},\ and\ \bibinfo
  {author} {\bibfnamefont {A.~F.}\ \bibnamefont {Kemper}},\ }\bibfield  {title}
  {\bibinfo {title} {{Fixed Depth {Hamiltonian} Simulation via {Cartan}
  Decomposition}},\ }\href {https://doi.org/10.1103/PhysRevLett.129.070501}
  {\bibfield  {journal} {\bibinfo  {journal} {Phys. Rev. Lett.}\ }\textbf
  {\bibinfo {volume} {129}},\ \bibinfo {pages} {070501} (\bibinfo {year}
  {2022})}\BibitemShut {NoStop}%
\bibitem [{\citenamefont {Wiebe}\ and\ \citenamefont
  {Granade}(2016)}]{Wiebe2016}%
  \BibitemOpen
  \bibfield  {author} {\bibinfo {author} {\bibfnamefont {N.}~\bibnamefont
  {Wiebe}}\ and\ \bibinfo {author} {\bibfnamefont {C.}~\bibnamefont
  {Granade}},\ }\bibfield  {title} {\bibinfo {title} {{Efficient {Bayesian}
  Phase Estimation}},\ }\href {https://doi.org/10.1103/PhysRevLett.117.010503}
  {\bibfield  {journal} {\bibinfo  {journal} {Phys. Rev. Lett.}\ }\textbf
  {\bibinfo {volume} {117}},\ \bibinfo {pages} {010503} (\bibinfo {year}
  {2016})}\BibitemShut {NoStop}%
\bibitem [{\citenamefont {Lee}\ \emph {et~al.}(2022)\citenamefont {Lee},
  \citenamefont {Qin}, \citenamefont {Raussendorf}, \citenamefont {Sela},\ and\
  \citenamefont {Scarola}}]{Lee2022}%
  \BibitemOpen
  \bibfield  {author} {\bibinfo {author} {\bibfnamefont {W.-R.}\ \bibnamefont
  {Lee}}, \bibinfo {author} {\bibfnamefont {Z.}~\bibnamefont {Qin}}, \bibinfo
  {author} {\bibfnamefont {R.}~\bibnamefont {Raussendorf}}, \bibinfo {author}
  {\bibfnamefont {E.}~\bibnamefont {Sela}},\ and\ \bibinfo {author}
  {\bibfnamefont {V.~W.}\ \bibnamefont {Scarola}},\ }\bibfield  {title}
  {\bibinfo {title} {Measurement-based time evolution for quantum simulation of
  fermionic systems},\ }\href
  {https://doi.org/10.1103/PhysRevResearch.4.L032013} {\bibfield  {journal}
  {\bibinfo  {journal} {Phys. Rev. Res.}\ }\textbf {\bibinfo {volume} {4}},\
  \bibinfo {pages} {L032013} (\bibinfo {year} {2022})}\BibitemShut {NoStop}%
\bibitem [{\citenamefont {Jordan}\ and\ \citenamefont
  {Wigner}(1928)}]{Jordan1928}%
  \BibitemOpen
  \bibfield  {author} {\bibinfo {author} {\bibfnamefont {P.}~\bibnamefont
  {Jordan}}\ and\ \bibinfo {author} {\bibfnamefont {E.}~\bibnamefont
  {Wigner}},\ }\bibfield  {title} {\bibinfo {title} {\"{U}ber das {P}aulische
  \"{A}quivalenzverbot},\ }\href {https://doi.org/10.1007/BF01331938}
  {\bibfield  {journal} {\bibinfo  {journal} {Z. Phys.}\ }\textbf {\bibinfo
  {volume} {47}},\ \bibinfo {pages} {631} (\bibinfo {year} {1928})}\BibitemShut
  {NoStop}%
\end{thebibliography}%

\end{document}